\documentclass[aps,prb,twocolumn,floatfix,nofootinbib]{revtex4-2}

\usepackage{graphicx}
\usepackage{multirow}
\usepackage{color}
\usepackage{amsmath}
\usepackage{amsfonts}
\usepackage{amssymb}
\usepackage{graphicx}
\graphicspath{{Figures/}}
\usepackage{epstopdf}
\usepackage{empheq}
\usepackage{float}
\usepackage{cases}
\usepackage{mathtools}
\usepackage{dcolumn}
\usepackage{physics}
\usepackage{bbold}
\usepackage{bm}
\usepackage[T1]{fontenc}
\usepackage{gensymb}

\usepackage{xurl} 
\usepackage[dvipsnames]{xcolor}
\usepackage[colorlinks]{hyperref}
\hypersetup{
colorlinks=True,
linkbordercolor=White,
linkcolor=BrickRed,
citecolor=Blue,	 
}

\renewcommand{\vec}[1]{\boldsymbol{#1}}

\begin{document}

\title{Modeling the CNOT gate in Germanium double quantum dots}

\author{Mauricio J. Rodr\'iguez}
\affiliation{Univ. Grenoble Alpes, CEA, IRIG-MEM-L\_Sim, Grenoble, France.}%
\author{Yann-Michel Niquet}
\email{yniquet@cea.fr}
\affiliation{Univ. Grenoble Alpes, CEA, IRIG-MEM-L\_Sim, Grenoble, France.}%

\date{\today}

\begin{abstract}
We model a controlled-NOT (CNOT) gate mediated by exchange interactions in a realistic Ge/GeSi hole spin-qubit device. We use for that purpose a computationally efficient framework based on a compact dressed basis of many-body singlet and triplet states, which enables fast time-dependent simulations that would be prohibitive within a conventional configuration-interaction approach. We discuss the dominant sources of infidelity in ideal (noise-free) devices. On the one hand, the nonadiabaticity of the tunnel barrier ramps, enhanced by the transverse component of the exchange interaction, imposes limits on the speed of the CNOT gate. On the other hand, the AC Stark shift of the conditional resonance and the off-resonant driving of the complementary transition can significantly degrade the gate fidelity and call for a careful optimization of the driving conditions. Finally, we investigate the impact of quasi-static charge noise on the fidelity of the CNOT gate. We find that the capture of charges near the qubits can strongly disrupt the conditional operation, while the nanometer-scale motion of charges results in a loss of fidelity that sensitively depends on the position and strength of the corresponding dipole. Our results quantify the requirements on barrier control, microwave driving, and charge-noise stability for high-fidelity two-qubit operations in Ge/GeSi hole-spin qubits.
\end{abstract}

\maketitle

\section{Introduction}
\label{sec:introduction}

Spin qubits \cite{Loss98,Burkard99} in semiconductor quantum dots \cite{Hanson07} have made remarkable progress in the past few years across several material platforms \cite{Stano22,Burkard2023Review}. High-fidelity one- and two-qubit gates have been demonstrated in electron spin qubits based on Si/SiO$_2$ and Si/SiGe heterostructures \cite{Yoneda18,Zajac18,Philips22,Mills22,Xue22,Noiri22,Huang24,Steinacker25,Wu26}, and cryo-electronic control of an array comprising 54 quantum dots (18 exchange-only qubits) has been recently achieved in Si/SiGe \cite{Abraham26}. Coherent spin shuttling between distant spin qubits has also been demonstrated in Si/SiGe \cite{DeSmet25,Undseth26}, opening up a promising path for the scaling of quantum dot arrays \cite{Kunne24}. In parallel, hole spin qubits have emerged as a compelling alternative to electron spin qubits \cite{Maurand16,Watzinger18,Hendrickx20b,Camenzind22,Fang2023Review}. Intrinsic spin-orbit coupling in the valence band \cite{Winkler03} indeed enables efficient electrical spin manipulation, at the possible cost of a larger (but manageable) sensitivity to electrical perturbations \cite{Bosco21,Froning21,Piot22,Mauro24,Hendrickx2024,Ungerer25,Bassi26}. Strained Ge/GeSi heterostructures have emerged as the leading platform for hole spin qubits \cite{Sammak19,Scappucci20}, with demonstrations of a variety of two-qubit gates (CROT, CPHASE, iSWAP...) \cite{Hendrickx20,Hendrickx21,Wang2024,Massai26}, as well as the operation of arrays containing up to 18 qubits \cite{Valentin25,Dijkema26}.

Exchange interactions arising from the competition between inter-dot tunneling and Coulomb repulsion \cite{Loss98,Burkard99,Hetenyi20,Geyer24} are the key ingredient of most two-qubit gates demonstrated to date. While single-qubit fidelities greater than $99.9$\% are now routinely reported for both electron and hole spin qubits \cite{Lawrie23,Wu26,Dijkema26}, high-fidelity two-qubit gates remain considerably more challenging \cite{Russ18,Heinz22,Russ23,CalderonVargas19,Russ2026}. As a matter of fact, the physics of coupled qubits is substantially richer and less understood than the physics of isolated qubits. Reliable two-qubit gates are, nevertheless, essential for quantum error correction and the scaling of quantum processors. From a theoretical point of view, the modeling of such two-qubit gates in realistic device layouts is particularly demanding. It requires many-body approaches such as configuration interaction (CI) methods \cite{Sherrill99,Rontani06,ElectroConfinedCI(EM),Abadillo21,e-eCI(TB),ValleytronicCI(EM),Valentin25,Secchi25,Bouquet2026}, whose computational cost can make extensive time-dependent calculations prohibitive. The CI method indeed expands the many-body wave functions in a basis of Slater determinants built from single-particle states. Individual Slater determinants do not, however, capture Coulomb correlations, which only emerge from the (slowly converging) mixing of such determinants. We have recently introduced a ``dressed basis'' approach \cite{MRodriguez2025} that ultimately replaces the large CI basis set by a compact set of singlet and triplet states sampled over a grid of relevant bias points. Because these states already incorporate the main Coulomb correlations, the resulting effective Hamiltonians enable fast and accurate time-dependent simulations while retaining the essential many-body physics.

In this work, we apply this framework to the simulation of a controlled-NOT (CNOT) gate between two hole spin qubits in a Ge/GeSi heterostructure \cite{Hendrickx20,Hendrickx21}. The CNOT operation flips the spin of the ``right'' qubit only if the spin in the ``left'' qubit is down. The protocol \cite{Zajac18,Russ18} consists in {\it i}) opening the tunnel barrier between the dots to turn on the exchange interaction and make the Larmor frequency $f_{\Uparrow/\Downarrow}$ of the right dot depend on the spin $\Uparrow/\Downarrow$ of the left dot, {\it ii}) driving resonantly at frequency $f_\Downarrow$ to flip the spin of the right dot only if the left dot is down, and {\it iii}) close the tunnel barrier and perform two single-qubit phase gates to correct for the extra phase accumulated during this sequence. We first analyze the sources of infidelity in the pristine (undisordered) device \cite{Russ18}. We find that the non-adiabaticity of the barrier ramps, enhanced by the so-called transverse exchange interaction, is a major limitation that calls for slow action on the tunnel barrier especially when the exchange is large. Besides, the AC Stark shift of the $f_{\Downarrow}$ transition and the non-intentional, off-resonant driving of the $f_{\Uparrow}$ transition can complicate the calibration of the CNOT gate and degrade the fidelity. The former can be mitigated by an appropriate choice of driving gate, and the second by a careful optimization of the drive power. We next investigate the robustness of the CNOT gate against quasi-static charge noise, modeled as a charge or dipole slowly fluctuating at the interface between the Ge/GeSi heterostructure and the gate stack. We find that the capture of a charge in the vicinity of the qubits disrupts the CNOT operation, whereas the motion of a charge $q$ over short distances $d\gtrsim 1$\,nm degrades the fidelity according to its exact location and dipole $p=qd$. This sets quantitative constraints on the level and spatial characteristics of charge noise compatible with high-fidelity two-qubit gates.

The paper is organized as follows: we review the methodology and the operation of CNOT gates in Section~\ref{sec:methodology}, then discuss the mechanisms limiting the fidelity in a pristine Ge/GeSi heterostructure in Section~\ref{sec:CNOTpristine}, and finally address the effects of quasi-static charge noise in Section~\ref{sec:noisyCNOT}.

\section{Methodology}
\label{sec:methodology}

\subsection{Device and numerical framework}
\label{sec:device}

\begin{figure}[t]
	\centering
 	\includegraphics[width=\linewidth]{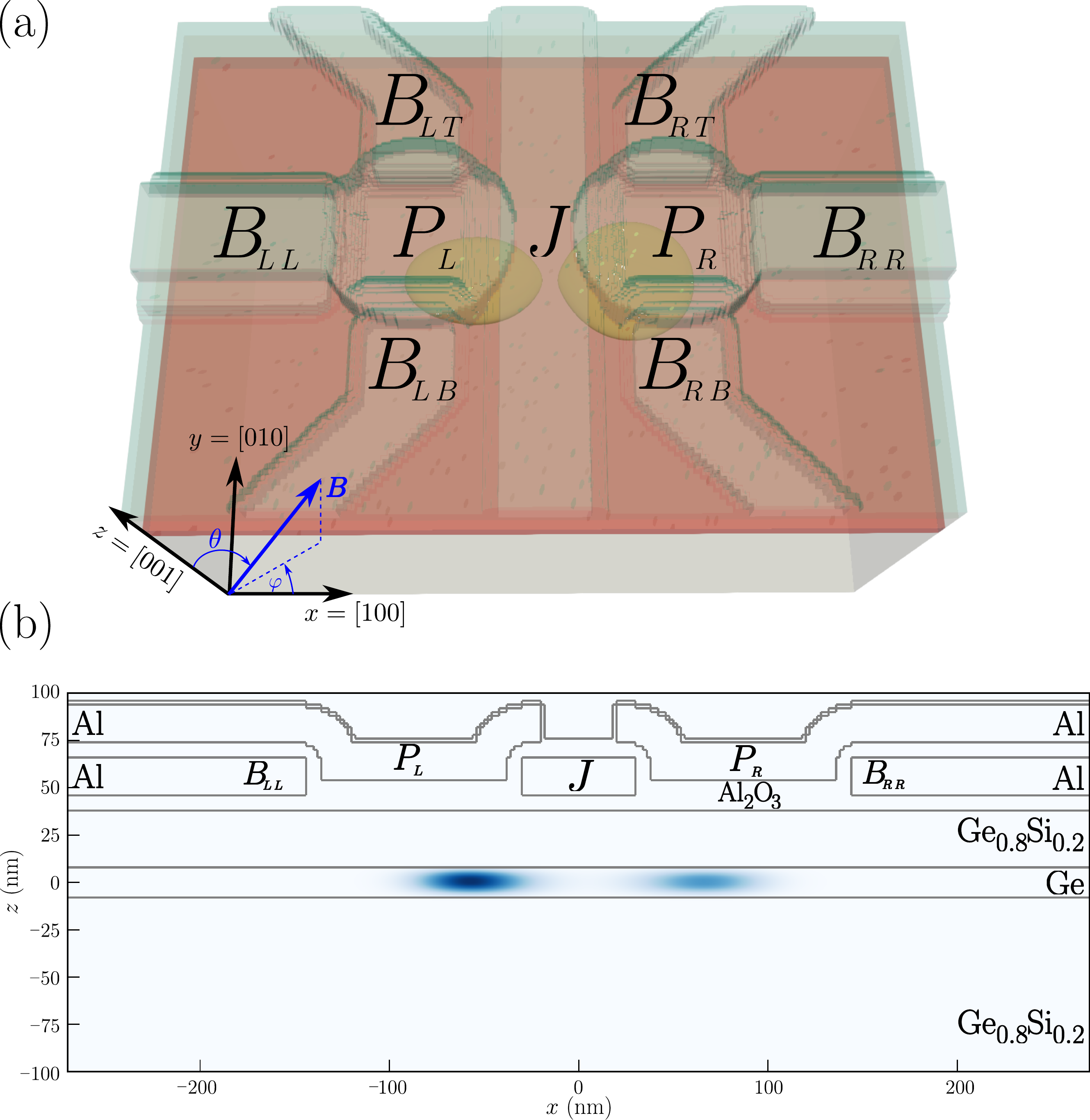}
	\caption{(a) The DQD device. The 16-nm-thick Ge well is colored in red, and Ge$_{0.8}$Si$_{0.2}$ in blue. The 20-nm-thick aluminium gates (in gray) are arranged on two levels separated from each other and from the heterostructure by 8\,nm of Al$_2$O$_3$: the $J$ gate and the barrier gates $B_{XY}$ on the first level, and the plunger gates $P_L$ and $P_R$ on the second. The diameter of the plunger gates is $d=130$\,nm and their centers are separated by $s=170$\,nm. The width of the $J$ gate is $w_J=60$\,nm and the width of the barrier gates is $w_B=90$\,nm. The orientation of the magnetic field $\vec{B}$ is characterized by the polar and azimuthal angles $\theta$ and $\varphi$. The yellow shape is the iso-density surfaces enclosing 85\% of the charge of the ground two-hole state at  $V_L=-72.5$\,mV, $V_R=-35$\,mV, $V_J=-15$\,mV, $V_{LT}=V_{LB}=40$\,mV, $V_{RT}=V_{RB}=-10$\,mV and $V_{LL}=V_{RR}=0$. (b) Cut in the $(xz)$ symmetry plane showing the two levels of gates, with a color map of the ground-state density (arbitrary units).}
	\label{fig:device}
\end{figure}

We consider the prototypical double-quantum-dot (DQD) device of Fig.~\ref{fig:device}. The heterostructure is a 16-nm-thick Ge quantum well strained by a Ge$_{0.8}$Si$_{0.2}$ buffer and capped with a Ge$_{0.8}$Si$_{0.2}$ barrier with thickness $t_\mathrm{GeSi}=30$\,nm. The biaxial strains in the well are, therefore, $\varepsilon_{xx}=\varepsilon_{yy}=-0.61\%$ and $\varepsilon_{zz}=0.45\%$ \cite{Mauro25}. The left ($L$) and right ($R$) dots are shaped by the difference of potential between two plunger gates $P_L$ and $P_R$ and side (barrier) gates $B_{XY}$. The tunnel coupling between the dots is controlled by the $J$ gate. 

We calculate the strains in the heterostructure with a finite-differences solver of the continuum elasticity equations, and the potential created by the gates with a finite-volumes solver of Poisson's equation \cite{martinez2022hole,Abadillo2023,Mauro25}. We account for the inhomogeneous strains imprinted by the differential thermal contraction of the materials as discussed in Ref.~\cite{Abadillo2023}. We next compute $N$ single-hole wave functions in these strains and potential with a finite-differences implementation of the Luttinger-Kohn (LK) Hamiltonian \cite{Luttinger55,Luttinger56,Winkler03,KP09}. Finally, we expand the two-hole wave functions of the DQD in the basis of the $M=N(N-1)/2$ Slater determinants built from these single-particle wave functions \cite{Abadillo21,MRodriguez2025}. As discussed in Ref.~\cite{MRodriguez2025}, this configuration interaction (CI) basis set is too large ($M\approx 5000$ if $N\approx 100$) for efficient time-dependent calculations. We build, therefore, a minimal, ``dressed'' basis set of $M'\ll M$ two-particle states by sampling the ground singlet and triplet CI wave functions over a grid of relevant bias points. The relevant quantities converge pretty fast as these states already capture the main Coulomb correlations. The time-dependent Schr\"odinger equation is then solved in the dressed basis set \cite{MRodriguez2025}. We consider hole with positive dispersion (see Appendix \ref{app:TDSELK} for details about the solution of the time-dependent Schr\"odinger equation for holes).

\subsection{Electronic structure of the DQD}

\begin{figure}[t]
	\centering
 	\includegraphics[width=.8\linewidth]{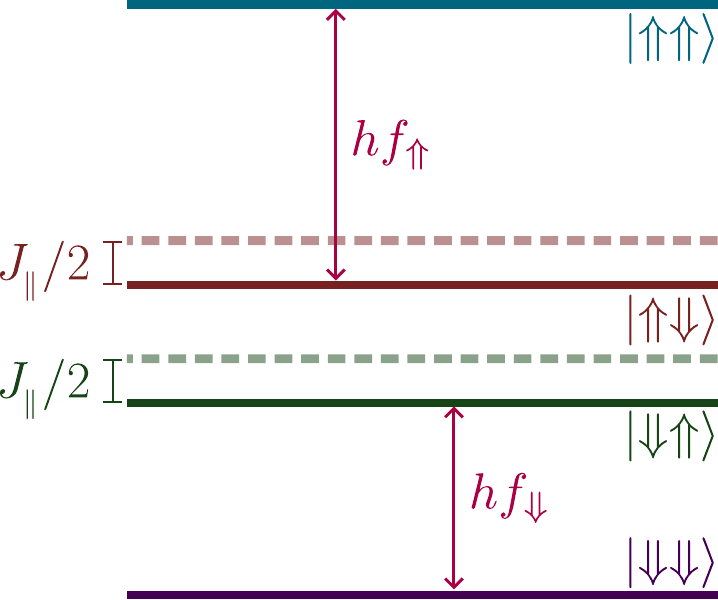}
	\caption{Scheme of the energy levels of an asymmetric two-hole DQD at high magnetic field. The dashed lines are the energies of the $\ket{\Uparrow\Downarrow}$ and $\ket{\Downarrow\Uparrow}$ states in the non-interacting limit. We also mark the relevant transition frequencies $f_\Uparrow=(E_{\Uparrow\Uparrow}-E_{\Uparrow\Downarrow})/h$ and $f_\Uparrow=(E_{\Downarrow\Uparrow}-E_{\Downarrow\Downarrow})/h$ that depend on the longitudinal exchange energy $J_\parallel$.}
	\label{fig:levels}
\end{figure}

We briefly review the electronic structure of the DQD in the $(1,1)$ charge state where both dots host one hole. At zero magnetic field, the two-hole ground-state is a singlet at energy $E_S=-J_0$ split from the nearby triplet at energy $E_T=0$ by the exchange energy $J_0$ \cite{Loss98,Burkard99,Burkard2023Review}. This exchange interaction primarily results from the interplay between inter-dot tunneling and Coulomb repulsion. The effects of a finite magnetic field $\vec{B}$ are twofold. On the one hand, $\vec{B}$ splits the triplet into $T_-$, $T_0$ and $T_+$ states with energies $-\bar{E}_Z$, $0$ and $+\bar{E}_Z$, where $\bar{E}_Z=\mu_B(g_L^*+g_R^*)B/2$ is the average Zeeman energy of the two dots ($\mu_B$ being Bohr's magneton and $g_{L,R}^*$ the effective $g$-factors of the $L$ and $R$ dots). On the other hand, $\vec{B}$ tends to break (mix) the singlet and $T_0$ states if $g_L^*\ne g_R^*$, as the two-hole eigenstates are the (anti-symmetrized) $\ket{\Uparrow\Downarrow}$ and $\ket{\Downarrow\Uparrow}$ states when $B\to\infty$ (where $\Uparrow$, $\Downarrow$ label the Zeeman-split states of each dot). This physics is described by the following effective Hamiltonian in the $\{\ket{\Uparrow\Uparrow},\ket{\Uparrow\Downarrow},\ket{\Downarrow\Uparrow},\ket{\Downarrow\Downarrow}\}$ basis set \cite{Geyer24}:
\begin{equation}
    H_e=\begin{pmatrix}
        \bar{E}_Z & 0 & 0 & 0\\
        0 & -\frac{1}{2}J_\parallel+\frac{1}{2}\Delta E_Z & \frac{1}{2}J_\perp & 0\\
        0 & \frac{1}{2}J_\perp & -\frac{1}{2}J_\parallel-\frac{1}{2}\Delta E_Z & 0\\
        0 & 0 & 0 & -\bar{E}_Z
    \end{pmatrix}\,,
\label{eq:Heff}
\end{equation}
where $\Delta E_Z=\mu_B(g_L^*-g_R^*)B$ is the difference between the Zeeman energies of the dots and $J_\parallel$, $J_\perp$ are respectively the longitudinal and transverse exchange energies. At high field $\Delta E_Z\gg J_\perp$, the eigenenergies of $H_e$ are indeed (see Fig.~\ref{fig:levels}):
\begin{subequations}
\label{eq:Es}
\begin{align}
    E_{\Downarrow\Downarrow}&=E_{T^-}=-\bar{E}_Z \\
    E_{\Downarrow\Uparrow}&\approx -\tfrac{1}{2}J_\parallel-\tfrac{1}{2}\Delta E_Z \\
    E_{\Uparrow\Downarrow}&\approx -\tfrac{1}{2}J_\parallel+\tfrac{1}{2}\Delta E_Z \\
    E_{\Uparrow\Uparrow}&=E_{T^+}=+\bar{E}_Z\,.
\end{align}
\end{subequations}
The longitudinal exchange thus splits the energies in one dot according to the spin in the other [e.g.,  $(E_{\Uparrow\Uparrow}-E_{\Uparrow\Downarrow})-(E_{\Downarrow\Uparrow}-E_{\Downarrow\Downarrow})=J_\parallel$]. At moderate magnetic fields, the transverse exchange mixes $\ket{\Uparrow\Downarrow}$ and $\ket{\Downarrow\Uparrow}$. The energies of the eigenstates $\{\ket{0},\ket{1},\ket{2},\ket{3}\}$ of $H_e$ are then:
\begin{subequations}
\label{eq:Es2}
\begin{align}
    E_0&=E_{T^-}=-\bar{E}_Z \\
    E_1&=-\tfrac{1}{2}J_\parallel-\tfrac{1}{2}\sqrt{\Delta E_Z^2+J_\perp^2} \\
    E_2&=-\tfrac{1}{2}J_\parallel+\tfrac{1}{2}\sqrt{\Delta E_Z^2+J_\perp^2} \\
    E_3&=E_{T^+}=+\bar{E}_Z\,.
\end{align}
\end{subequations}
We emphasize that $J_\parallel$ and $J_\perp$ are both different from the exchange energy $J_0$ at zero magnetic field due to spin-orbit coupling (SOC) \cite{Hetenyi20,Geyer24,Saezmollejo25}. Close to the so-called symmetric operation point (where the two dots are tuned), and in the simplest Hund-Mulliken model \cite{Loss98,Burkard99} (see Appendix \ref{app:HundMulliken}),
\begin{subequations}
\label{eq:Js}
\begin{align}
    J_0&\approx\frac{4t^2}{U} \\
    J_\parallel&\approx J_0-\frac{8t_\mathrm{sf}^2}{U} \\
    J_\perp&\approx J_0-\frac{4t_\mathrm{sf}^2}{U}\,,
\end{align}
\end{subequations}
where $U$ is the charging energy (assumed equal for both dots for simplicity), and $t=\sqrt{t_\mathrm{sc}^2+t_\mathrm{sf}^2}$ is the total tunnel coupling between the dots, split into a spin-conserving component $t_\mathrm{sc}$ (coupling the $\ket{\Uparrow}$ states or the $\ket{\Downarrow}$ states of each dot), and a spin-flip component $t_\mathrm{sf}$ (coupling the $\ket{\Uparrow}$ state of one dot to the $\ket{\Downarrow}$ state of the other). The latter collects contributions from different fingerprints of SOC in the DQD, such as the mismatch between the $g$-tensors of the isolated dots and the Rashba/Dresselhaus interactions undergone while tunneling \cite{Stepanenko12,froning_strong_SO_2021,jirovec_Freqs_ST-Ge,mutter_all-electrical_ST-Ge,yu2022strong,Jirovec23,MRodriguez2025}. Both $t_\mathrm{sc}$ and $t_\mathrm{sf}$ (thus $J_\parallel$ and $J_\perp$) depend on the orientation of the magnetic field (as do the Zeeman-split $\ket{\Uparrow}$ and $\ket{\Downarrow}$ states). Since $t_\mathrm{sf}\le t$, $-J_0\le J_\parallel<J_0$ can be negative, $0\le J_\perp\le J_0$ and $J_\perp\approx(J_0+J_\parallel)/2$ \cite{Massai26}.

\subsection{The CNOT gate}

According to Eqs.~\eqref{eq:Es}, the high-field transitions $\ket{\Uparrow\Downarrow}\leftrightarrow\ket{\Uparrow\Uparrow}$ at frequency $f_\Uparrow=(E_{\Uparrow\Uparrow}-E_{\Uparrow\Downarrow})/h$ and $\ket{\Downarrow\Downarrow}\leftrightarrow\ket{\Downarrow\Uparrow}$ at frequency $f_\Downarrow=(E_{\Downarrow\Uparrow}-E_{\Downarrow\Downarrow})/h$ are detuned by the longitudinal exchange energy $J_\parallel$ (namely, $f_\Uparrow-f_\Downarrow=J_\parallel/h$). Therefore, sending a radio-frequency (RF) pulse with frequency $f_\Downarrow$ on a gate drives the $R$ dot spin only if the $L$ dot is in the $\ket{\Downarrow}$ state. A $\pi$-pulse thus ideally results in a CNOT operation \cite{Zajac18,Russ18}, defined by the following evolution matrix in the $\{\ket{\Uparrow\Uparrow},\ket{\Uparrow\Downarrow},\ket{\Downarrow\Uparrow},\ket{\Downarrow\Downarrow}\}$ basis set:
\begin{equation}
    U_0=\begin{pmatrix}
		1 & 0 & 0 & 0 \\
		0 & 1 & 0 & 0 \\
		0 & 0 & 0 & 1 \\
		0 & 0 & 1 & 0
	\end{pmatrix}.
    \label{eq:U0}
\end{equation}
We shall, nevertheless, account for various phase corrections and distinguish the computational basis set (barrier closed) from the operational basis set (barrier open) where the CNOT is driven.

\subsubsection{Computational and operational subspaces}

The exchange interactions in a quantum register are expected to be tunable enough to be switched ``on'' and ``off'' on demand. In particular, the initial (preparation) and final (measurement) states of the two qubits are usually defined in a ``computational'' subspace with exchange off (barrier closed, isolated qubits), while the conditional $\pi$-pulse of the CNOT gate is performed in an ``operational'' subspace with exchange on (barrier open, interacting qubits). 

We therefore distinguish the computational basis $\{\ket{\Uparrow\Uparrow},\ket{\Uparrow\Downarrow},\ket{\Downarrow\Uparrow},\ket{\Downarrow\Downarrow}\}$ with barrier closed ($J_\parallel\approx J_\perp\approx 0$) from the operational basis $\{\ket{3},\ket{2},\ket{1},\ket{0}\}$ with barrier open (the eigenstates of $H_e$ at finite $J_\perp$), where $\ket{1}$ and $\ket{2}$ generally mix $\ket{\Uparrow\Downarrow}$ and $\ket{\Downarrow\Uparrow}$ at reasonable magnetic fields. This does not preclude the CNOT operation, as $f_\Uparrow=(E_3-E_2)/h$ and $f_\Downarrow=(E_1-E_0)/h$ remain split by $f_\Uparrow-f_\Downarrow\approx J_\parallel/h$ [see Eqs.~\eqref{eq:Es2}]. The barrier must thus be adiabatically opened to map the basis states of the computational subspace onto those of the operational subspace, then adiabatically closed after the $\pi$-pulse to bring the result back to the computational subspace. We will discuss the role of the transverse exchange interaction on this mapping in section \ref{sec:Jramps}. We can then write down the $4\times 4$ evolution matrix $U_{ij}=\braket{\psi_i(0)}{\psi_j(\tau_\mathrm{CNOT})}$ that characterizes the gate in the computational basis set [$\ket{\psi_i(0)}$ being an initial state among $\{\ket{\Uparrow\Uparrow},\ket{\Uparrow\Downarrow},\ket{\Downarrow\Uparrow},\ket{\Downarrow\Downarrow}\}$ and $\ket{\psi_i(\tau_\mathrm{CNOT})}$ the corresponding final state at time $t=\tau_\mathrm{CNOT}$]. Note that $U$ may not be unitary if the final states leak outside the computational subspace.

\subsubsection{Phase corrections}

On top of the $\pi$-pulse, the different states may dephase as their energies can be modulated throughout the operation. This is best characterized in the co-rotating frame of the computational subspace, by the matrix $\hat{U}=R(\tau_\mathrm{CNOT})U$, where $\tau_\mathrm{CNOT}$ is the total gate time and $R(t)$ the transformation:
\begin{equation}
    R(t)=\begin{pmatrix}
 		e^{i\omega_+t} & 0 & 0 & 0 \\
 		0 & e^{i\omega_-t} & 0 & 0 \\
 		0 & 0 & e^{-i\omega_-t} & 0 \\
 		0 & 0 & 0 & e^{-i\omega_+t} 
 	\end{pmatrix}\,.
\end{equation}
Here $\omega_+=\bar{E}_Z/\hbar=\mu_B(g_L^*+g_R^*)B/(2\hbar)$, and $\omega_-=\Delta E_Z/(2\hbar)=\mu_B(g_L^*-g_R^*)B/(2\hbar)$ are computed barrier closed. The free evolution operator is thus the identity in the co-rotating frame, so that $\hat{U}$ captures the drifts in phase with respect to that free evolution.

From now on, we characterize all operations in the co-rotating frame and drop the hat on the evolution matrices for simplicity. Ideal $\pi$-pulse and barrier ramps are generally expected to yield the operation:
\begin{equation}
    U_\pi=\begin{pmatrix}
 		1 & 0 & 0 & 0 \\
 		0 & e^{i\delta\varphi_2} & 0 & 0 \\
 		0 & 0 & 0 & e^{i\delta\varphi_0} \\
 		0 & 0 & e^{i\delta\varphi_1} & 0
	\end{pmatrix}\,,
	\label{eq:U}
\end{equation}
where we have factorized an (irrelevant) global phase factor, and the $\delta\varphi_i$ (defined modulo $2\pi$) characterize the relative phases accumulated by the computational states during the CNOT gate. These phases must be corrected to match $U$ to the target $U_0$ [Eq.~\eqref{eq:U0}]. For that purpose, we can perform three independent $Z$-rotations on the left and right dots (e.g., one rotation on the $L$ dot before opening the barrier, and one rotation on the $L$ and one rotation on the $R$ dot after closing the barrier). Such $Z$-rotations can in principle be achieved by applying a square voltage pulse on one or more gates in order to detune the Larmor frequencies of the left or right qubit for a time $\tau$ (phase gates). The matrix $Z(\tau)$ of this operation reads in the co-rotating frame:
\begin{equation}
    Z(\tau)=\begin{pmatrix}
        e^{-i\delta\omega_+\tau} & 0 & 0 & 0 \\
        0 & e^{-i\delta\omega_-\tau} & 0 & 0 \\
        0 & 0 & e^{i\delta\omega_-\tau} & 0 \\
        0 & 0 & 0 & e^{i\delta\omega_+\tau} \\
    \end{pmatrix}\,,
\end{equation}
where $\delta\omega_\pm=(\delta\omega_L\pm\delta\omega_R)/2$ and $\delta\omega_{L/R}=\mu_B\delta g_{L/R}^*B/\hbar$, with $\delta g_L^*$, $\delta g_R^*$ the variations of the effective $g$-factors of the left and right dots during the pulse. This is indeed the composition of one rotation of angle $-\delta\omega_L\tau$ around $Z$ in the $L$ dot, and one rotation of angle $-\delta\omega_R\tau$ around $Z$ in the $R$ dot. We assume below that the $\delta\omega_{L/R}$ are given (and set by the voltage pulse on the gates), and that the rotations are controlled by the pulse duration $\tau$.

Nevertheless, we can drop the first rotation by waiting a suitable time with the barrier open \cite{Russ18}. Let us indeed perform two phase gates $Z_1(\tau_1)$ and $Z_2(\tau_2)$ (with different $\delta\omega_L$ and $\delta\omega_R$) after closing the barrier. The resulting operation $U=Z_2(\tau_2)Z_1(\tau_1)U_\pi$ takes the form:
\begin{equation}
   \label{eq:Uphi}
    U=\begin{pmatrix}
 		1 & 0 & 0 & 0 \\
 		0 & e^{i(\delta\varphi_2-\phi_2)} & 0 & 0 \\
 		0 & 0 & 0 & e^{i(\delta\varphi_1-\phi_1)} \\
 		0 & 0 & e^{i(\delta\varphi_0-\phi_0)} & 0
	\end{pmatrix}\,,
\end{equation}
where:
\begin{subequations}
\label{eq:phis}
\begin{align}
    \phi_0&=-2(\delta\omega_{1+}\tau_1+\delta\omega_{2+}\tau_2) \\ \phi_1&=-(\delta\omega_{1+}+\delta\omega_{1-})\tau_1-(\delta\omega_{2+}+\delta\omega_{2-})\tau_2 \\ \phi_2&=-(\delta\omega_{1+}-\delta\omega_{1-})\tau_1-(\delta\omega_{2+}-\delta\omega_{2-})\tau_2\,.
\end{align}
\end{subequations}
Since $\phi_0=\phi_1+\phi_2$, the $\delta\varphi$ must fulfill $\delta\varphi_0=\delta\varphi_1+\delta\varphi_2\ [2\pi]$ in order to correct all three phases with only two independent phase gates. Yet if we wait for an additional time $\tau_w$ after the $\pi$ pulse before closing the barrier, the $\delta\varphi$ in Eqs.~\eqref{eq:U} and \eqref{eq:phis} become
\begin{subequations}
\begin{align}
   \delta\varphi_0^\prime&=\delta\varphi_0-(\omega_0-\omega_3)\tau_w \\
   \delta\varphi_1^\prime&=\delta\varphi_1-(\omega_1-\omega_3)\tau_w \\
   \delta\varphi_2^\prime&=\delta\varphi_2-(\omega_2-\omega_3)\tau_w\,,
\end{align}
\end{subequations}
where $\omega_i=E_i/\hbar$ and the $E_i$ are the energies of the operational states  [Eq.~\eqref{eq:Es2} with the barrier open], so that:
\begin{align}
\delta\varphi_0^\prime&-\delta\varphi_1^\prime-\delta\varphi_2^\prime \nonumber \\
&=\delta\varphi_0-\delta\varphi_1-\delta\varphi_2-(\omega_3-\omega_2-\omega_1+\omega_0)\tau_w \nonumber \\
&=\delta\varphi_0-\delta\varphi_1-\delta\varphi_2-J_\parallel\tau_w/\hbar\,.
\end{align}
We can hence choose $\tau_w$ such that $\delta\varphi_0^\prime=\delta\varphi_1^\prime+\delta\varphi_2^\prime\ [2\pi]$, and then solve the equations $\delta\varphi_i^\prime=\phi_i$, which yields:
\begin{subequations}
\label{eq:rottimes}
\begin{align}
	\tau_1&=\frac{-\left(\delta\varphi_1^\prime+2n\pi\right)\delta\omega_{2R}+\left(\delta\varphi_2^\prime+2m\pi\right)\delta\omega_{2L}}{\delta\omega_{1L}\delta\omega_{2R}-\delta\omega_{1R}\delta\omega_{2L}} \\
	\tau_2&=\frac{+\left(\delta\varphi_1^\prime+2n\pi\right)\delta\omega_{1R}-\left(\delta\varphi_2^\prime+2m\pi\right)\delta\omega_{1L}}{\delta\omega_{1L}\delta\omega_{2R}-\delta\omega_{1R}\delta\omega_{2L}} \,,
\end{align}
\end{subequations}
where $n$ and $m$ are relative integers (as the $\delta\varphi^\prime$ are defined modulo $2\pi$). There is thus an infinity of solutions, the optimal one minimizing $\tau_1+\tau_2$. We emphasize that separate rotations on the $L$ and $R$ dots are not a pre-requisite for phase corrections; the $(\delta\omega_L,\delta\omega_R)$ vectors of the two phase gates must, however, be as orthogonal as possible for best efficiency\footnote{There are actually no solutions if they are parallel, as the two phase gates commute.}.

The experimental calibration of the CNOT gate has been discussed in Ref.~\cite{Zajac18}. In the numerical calculations, we tune $\tau_w$ so that the relevant matrix elements of $U$ satisfy $\delta\varphi_0=\delta\varphi_1+\delta\varphi_2\ [2\pi]$ after closing the barrier, and then adjust $\tau_1$ and $\tau_2$ to maximize the fidelity (defined below) using Eq.~\eqref{eq:rottimes} as a first guess (see section \ref{sec:demonstration}).

\subsubsection{Fidelity}

In the following, we characterize the performance of the CNOT gate by the fidelity \cite{Pedersen07b}
\begin{equation}
    \mathcal{F}=\frac{1}{n(n+1)}\left[\mathrm{Tr}\left(U^\dagger U\right)+\abs{\mathrm{Tr}\left(U_0^\dagger U\right)}^2\right]\,,
    \label{eq:Fidelity}
\end{equation}
where $n=4$ is the dimension of the computational subspace. The first term in the brackets characterizes leakage outside the computational subspace (how far $U$ is unitary), while the second term measures how close $U$ is to $U_0$ (but is also degraded by leakage).

\subsection{Choice of bias point and characterization of the device}

We bias the barrier gates of the device so that the $L$ and $R$ dots are different (as they usually are experimentally). Specifically, we ground $B_{LL}$ and $B_{RR}$ and set $V_{LT}=V_{LB}=40$\,mV and $V_{RT}=V_{RB}=-10$\,mV. This squeezes the left dot along $y$, as shown by the iso-density surfaces of the ground two-hole state plotted in Fig.~\ref{fig:device} at $V_L=-72.5$\,mV, $V_R=-35$\,mV and $V_J=-15$\,mV. This bias is actually close to the symmetric operation point where $\partial J_0/\partial\varepsilon=0$, with $\varepsilon$ the detuning energy between the $L$ and $R$ dots. The exchange energy is thus little sensitive to detuning noise at that point.

\begin{figure}[t]
	\centering
 	\includegraphics[width=\linewidth]{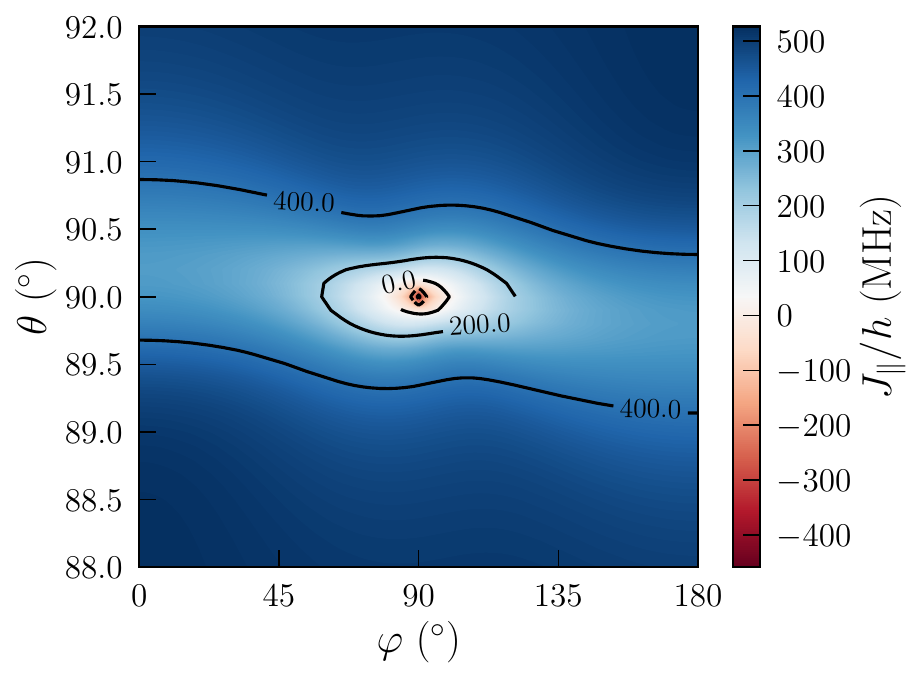}
	\caption{Map of the longitudinal exchange energy $J_\parallel$ as a function of the orientation of the magnetic field characterized by the angles $\theta$ and $\varphi$ defined in Fig.~\ref{fig:device}. The bias point is $V_L=-72.5$\,mV, $V_R=-35$\,mV and $V_J=-15$\,mV.}
	\label{fig:Jparamap}
\end{figure}

The longitudinal exchange energy at that bias is plotted as a function of the orientation of the magnetic field in Fig.~\ref{fig:Jparamap}. It is estimated as $J_\parallel=E_3-E_2-E_1+E_0$ along the lines of Ref.~\cite{MRodriguez2025}. It shows strong modulations around in-plane orientations due to the mismatch of the $g$-tensors of the two dots, the Rashba interactions being usually weak in strained Ge \cite{Marcellina17,Terrazos21,Bosco21b,Rodriguez2023,Saezmollejo25,MRodriguez2025}. Interestingly, $J_\parallel$ becomes negative when $\vec{B}$ comes close to the $y$ axis because the diagonal element $g_{yy}$ of these $g$-tensors has opposite sign in the two dots (due to the sizable squeezing of the $L$ dot). Therefore, the $\ket{\Uparrow}$ and $\ket{\Downarrow}$ states have opposite angular momentum components in the $L$ and $R$ dots; thus tunneling processes that conserve the {\it physical} spin appear as (pseudo-)spin-flip terms between $\ket{\Uparrow}$ and $\ket{\Downarrow}$ ($t_\mathrm{sc}\to0$, $t_\mathrm{sf}\to t$). Note that the line $J_\parallel=0$ is ideally suited to the implementation of SWAP gates \cite{Massai26}; however in the present study we set $\vec{B}\parallel\vec{x}$ in order to achieve both large $\abs{J_\parallel}$ and fast $\pi$ pulses (see later discussion\footnote{A magnetic field $\vec{B}\parallel\vec{y}$ does achieve both large $|J_\parallel|$ and small $|J_\perp|$ [see Eqs.~\eqref{eq:Js}], which is in principle desirable for a CNOT gate. This is, however, a pretty unconventional (and unstable) situation, as the the $g$-factor of the left dot is very small and highly sensitive (as is $J_\parallel$) to perturbations.}).

\begin{figure}[t]
	\centering
 	\includegraphics[width=\linewidth]{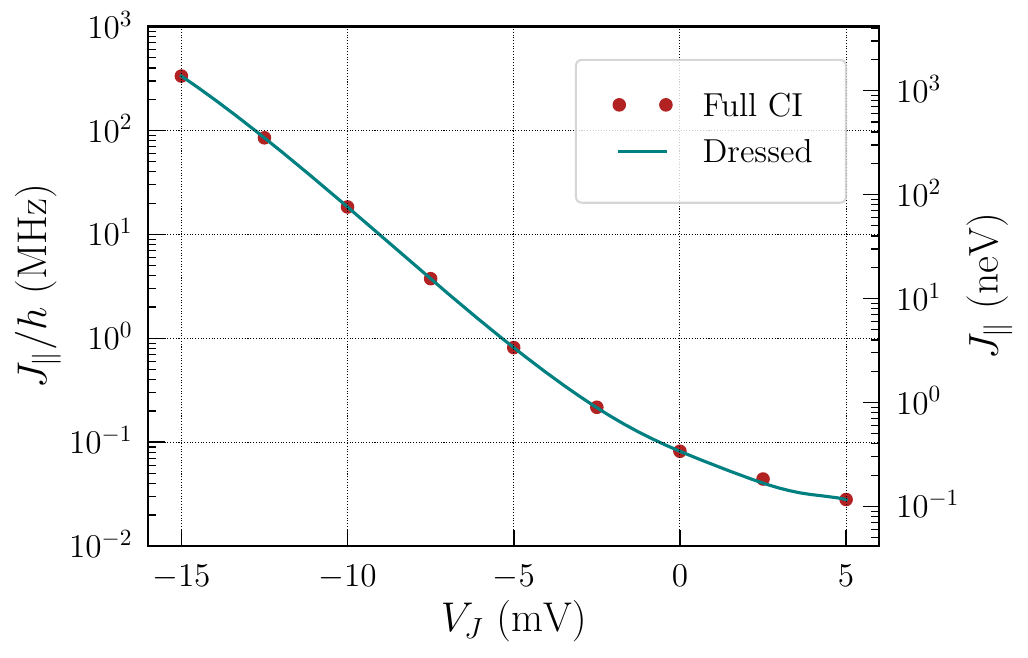}
	\caption{Longitudinal exchange energy $J_\parallel$ as a function of the $J$-gate voltage $V_J$. The bias is $V_L=-72.5$\,mV, $V_R=-35$\,mV, and the magnetic field is $B=1$\,T along $x$. The dressed basis set data (line) are compared to full CI calculations (dots).}
	\label{fig:JparaVJ}
\end{figure}

For this orientation, we estimate $J_\parallel/h=334$\,MHz at $V_J=-15$\,mV. In order to assess the performances of the CNOT gate, we also consider other $V_J$ with smaller exchange energies, listed in Table \ref{tab:exchange}. The longitudinal exchange energy is plotted as a function of $V_J$ in Fig.~\ref{fig:JparaVJ}. As expected, it decreases almost exponentially with increasing the $J$-gate voltage up to $V_J\approx 0$\,V. The trend is, however, much smoother when $V_J\gtrsim 0$\,V, and it looks difficult to reduce $J_\parallel$ down to the $0.1$ neV range. Although this range is hard to converge numerically, we emphasize that this saturation is physical. Due to SOC, the wave functions of the $\ket{\Uparrow}$ and $\ket{\Downarrow}$ states are (spatially) different in a finite magnetic field, which gives rise to a spin-dependent dipolar interaction\footnote{Although this interaction is purely transverse (XY) in Refs.~\cite{Flindt06,Trif07}, the $g$-tensors of the two dots are different in the present case \cite{MRodriguez2025}, which shall also give rise to a longitudinal ZZ component.} between the dots \cite{Flindt06,Trif07}. Although small, this electrostatic interaction prevails once the barrier is closed and the wave functions do not overlap any more. It sets a floor for the effective, residual interactions between the dots.

\begin{table}[t]
	\centering
	\begin{ruledtabular}
	    \begin{tabular}{lrrrr}
		    $V_J^\mathrm{op}$ (mV) & $-15.00$ & $-12.78$ & $-11.62$ & $-10.50$\\
		    $J_0/h$ (MHz) & $525$ & $143$ & $69$ & $33$\\
		    $J_\parallel/h$ (MHz) & $334$ & $100$ & $50$ & $25$\\
		    ${\bar J}_\perp/h$ (MHz) & $430$ & $122$ & $60$ & $29$\\
		    $J_\perp^*/h$ (MHz) & $411$ & $115$ & $55$ & $27$ 
		\end{tabular}
	\end{ruledtabular}
	\caption{Calculated $J_0$, $J_\parallel$ and estimated $J_\perp$ [as ${\bar J}_\perp\approx (J_0+J_\parallel)/2$ or as $J_\perp^*$ from Eq.~\eqref{eq:Jperp}] for different operational $J$-gate voltages $V_J^\mathrm{op}$.}
	\label{tab:exchange}	
\end{table}

We consider the barrier closed at $V_J=5$\,mV. The computational basis states are defined as the eigenstates of the CI Hamiltonian at that point. The $g$-factors of the dots are then $g_L^*=0.228$ and $g_R^*=0.125$. We set $B=1$\,T so that the Larmor frequencies of the $L$ and $R$ dots are $f_L=g_L^*\mu_BB/h=3.19$\,GHz and $f_R=g_R^*\mu_BB/h=1.75$\,GHz. The residual exchange energy at that bias and magnetic field is $J_\parallel=24$\,kHz.

\section{The CNOT gate in the pristine device}
\label{sec:CNOTpristine}

In this section, we demonstrate a CNOT gate in the device of Fig.~\ref{fig:device}, and analyze the fidelity of the various parts of the operation (exchange ramps, $\pi$-pulse, phase gates).

\subsection{Demonstration of the CNOT gate}
\label{sec:demonstration}

\begin{figure}[t]
	\centering
 	\includegraphics[width=\linewidth]{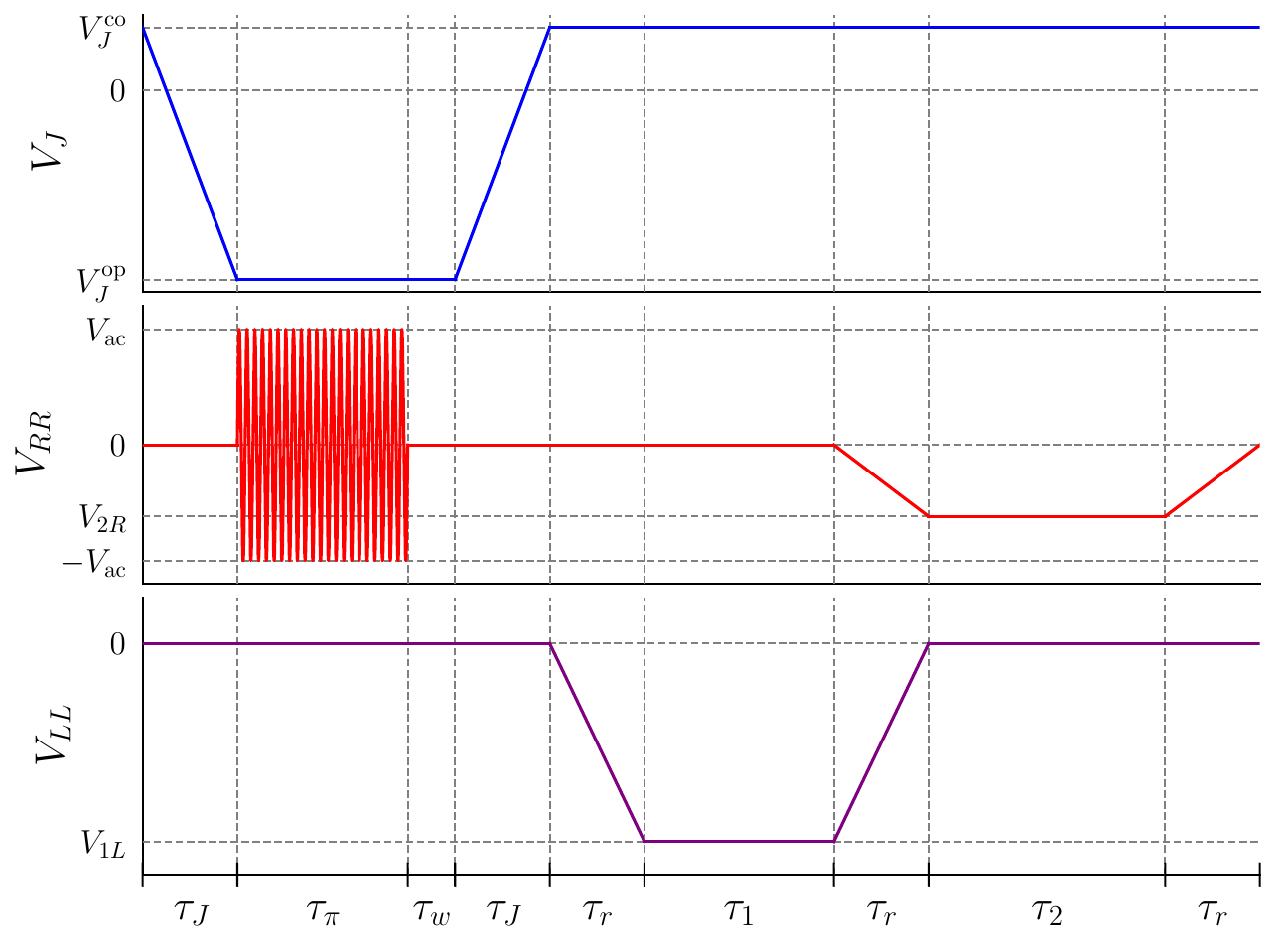}
	\caption{Timing diagram of the pulse sequence applied to the different gates in order to achieve the CNOT operation.}
	\label{fig:sequence}
\end{figure}

We start with the barrier closed at $V_J=5$\,mV ($V_L=-72.5$\,mV, $V_R=-35$\,mV). We first linearly ramp the $J$-gate voltage up to $V_J=-15$\,mV over a time $\tau_J$. We then drive the right qubit at frequency $f_\Downarrow$ for a time $\tau_\pi$ with the $B_{RR}$ gate (the choice of this gate will be justified in section \ref{sec:drive}). We next wait for a time $\tau_w$, and ramp down the $J$-gate voltage back to $V_J=5$\,mV (over the same time $\tau_J$). We finally perform two phase gates, with a pulse of $-1$\,mV on gate $B_{LL}$, then a pulse of $-1$\,mV on gate $B_{RR}$ (both with rise and fall times $\tau_r$). We illustrate this whole sequence on Fig.~\ref{fig:sequence}.

For the first phase gate ($V_{LL}=-1$\,mV), we get $\delta\omega_{1L}/(2\pi)=14.75$\,MHz and $\delta\omega_{1R}/(2\pi)=1.36$\,MHz. For the second phase gate ($V_{RR}=-1$\,mV), we get likewise $\delta\omega_{2L}/(2\pi)=1.36$\,MHz and $\delta\omega_{2R}/(2\pi)=25.46$\,MHz. As expected, the pulse on $B_{LL}$ results in a larger $Z$-rotation of the $L$ than of the $R$ qubit (and vice-versa for the pulse on $B_{RR}$). The $(\delta\omega_{1L},\delta\omega_{1R})$ and $(\delta\omega_{2L},\delta\omega_{2R})$ vectors are almost orthogonal (the two phase gates are reasonably independent), so that we did not optimize the phase gate pulses further.

With the barrier open at $V_J=-15$\,mV, $f_\Uparrow=2.755$\,GHz and $f_\Downarrow=2.421$\,GHz ($J_\parallel/h=f_\Uparrow-f_\Downarrow=334$\,MHz). We drive the $B_{RR}$ gate with a RF signal\footnote{More precisely, $V_{RR}=V_\mathrm{ac}\sin(2\pi f_\Downarrow (t-t_0))$ with $t_0$ the starting time of the drive. We also drive for $\tau_\pi=n/f_\Downarrow$ (with $n$ an integer) so that there is no discontinuity in the signals applied to the gates (only the slopes may be discontinuous). Therefore, $\tau_\pi$ may slightly over/undershoot the intended $\pi$ pulse. We did not attempt to perform further pulse optimization at this stage.} $V_{RR}=V_\mathrm{ac}\sin(2\pi f_\Downarrow t+\varphi)$. We set $V_\mathrm{ac}=1$\,mV, so that the Rabi frequency is $\nu=28.44$\,MHz and the drive time $\tau_\pi=1/(2\nu)=17.76$\,ns. As discussed in section \ref{sec:methodology}, we adjust $\tau_w$ so that the phases $\delta\varphi_0$, $\delta\varphi_1$ and $\delta\varphi_2$ of the relevant matrix elements of $U$ after the barrier is closed [Eq.~\eqref{eq:rottimes}] satisfy $\delta\varphi_0=\delta\varphi_1+\delta\varphi_2\ [2\pi]$ and can be corrected by the two phase gates.

\begin{figure}[t]
	\centering
 	\includegraphics[width=\linewidth]{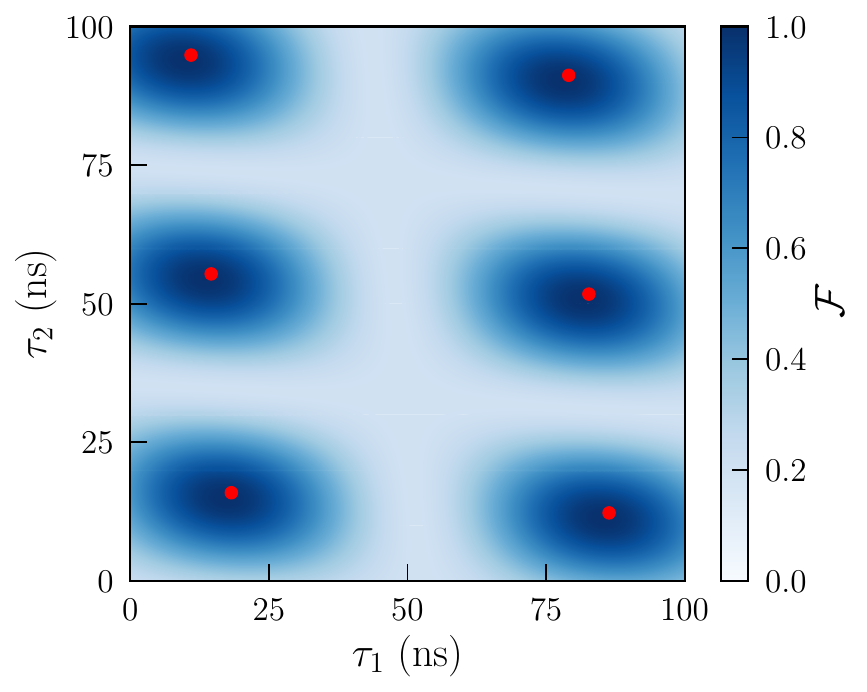}
	\caption{Map of the fidelity of the CNOT gate as a function of the times $\tau_1$ and $\tau_2$ of the two phase gates. The red dots are the optimal times expected from Eqs.~\eqref{eq:rottimes}. Here $\tau_J=10$\,ns, $\tau_r=1$\,ns, $\tau_\pi=17.76$\,ns ($V_\mathrm{ac}=1$\,mV), and $\tau_w=17.52$\,ns.}
	\label{fig:Ftau1tau2}
\end{figure}

We plot the fidelity of the resulting operation as a function of the phase gate times $\tau_1$ and $\tau_2$ in Fig.~\ref{fig:Ftau1tau2}. The ramp times are $\tau_J=10$\,ns for the $J$ gate and $\tau_r=1$\,ns for the phase gates (see optimization in section \ref{sec:Jramps}). The corresponding wait time is $\tau_w=17.52$\,ns. The fidelity exhibits the periodic pattern expected from Eqs.~\eqref{eq:rottimes} \cite{Russ18}. However, the position of the maxima is slightly shifted with respect to Eqs.~\eqref{eq:rottimes}, primarily because the latter do not account for the finite ramp time $\tau_r$ of the phase gates. For the shortest optimal $\tau_1=17.24$\,ns and $\tau_2=14.92$\,ns (total gate time $\tau_\mathrm{CNOT}=2\tau_J+\tau_\pi+\tau_w+3\tau_r+\tau_1+\tau_2=90.28$\,ns), the fidelity reaches $\mathcal{F}=99.90$\%, and is limited by the resolution of the map and the various errors (calibration, drive and diabaticity of the ramps) discussed below. This demonstrates an effective CNOT operation.

From a technical point of view, the time-dependent simulations are performed with $10\,000$ time steps in each ramp and $1\,000$ time steps per Larmor period $1/f_\Downarrow$ during the drive\footnote{The number of time steps during the wait time $\tau_w$ and phase gates is irrelevant as the control signals are stationary and the evolution is computed from the exact diagonalization of the Hamiltonian at each time step \cite{MRodriguez2025}.}. We compute $N=96$ single-hole states at $V_L=-72.5$\,mV, $V_R=-35$\,mV, $V_J=0$\,mV and we construct the dressed basis from the lowest-lying CI singlet and triplet states at the bias points $(V_J,\delta V,V_{LL},V_{RR})$ on the grid $(V_J,\delta V,0,0)$ with $V_J\in\{-15,-10,-5,0,5\}$\,mV and $\delta V\in\{-10.09,-5.05,0,5.05,10.09\}$\,mV (with $\delta V=\delta V_L-\delta V_R$ is the total detuning voltage between the dots). We also add the points $(0,0,-1,0)$\,mV, $(0,0,0,-1)$\,mV, $(5,0,-1,0)$\,mV, and $(5,0,0,-1)$\,mV as we use the $B_{LL}$ and $B_{RR}$ gates to drive the CNOT and for the phase gates. Since we work at a fixed $B=1$\,T along $x$, we directly include the magnetic field in the full CI calculations for better accuracy. The dimension of the dressed basis is $N=116$. The longitudinal exchange energies $J_\parallel$ computed in the CI and dressed basis sets, compared in Fig.~\ref{fig:JparaVJ}, are in excellent agreement over the whole range of $J$-gate voltages.

We now analyze the fidelity of the different parts of the CNOT sequence.

\subsection{Fidelity of the $J$-gate ramps}
\label{sec:Jramps}

We first address the fidelity of the $J$-gate ramps. We focus on the opening of the barrier, the situation being fairly symmetric when closing the barrier.

\begin{figure}[t]
	\centering
 	\includegraphics[width=\linewidth]{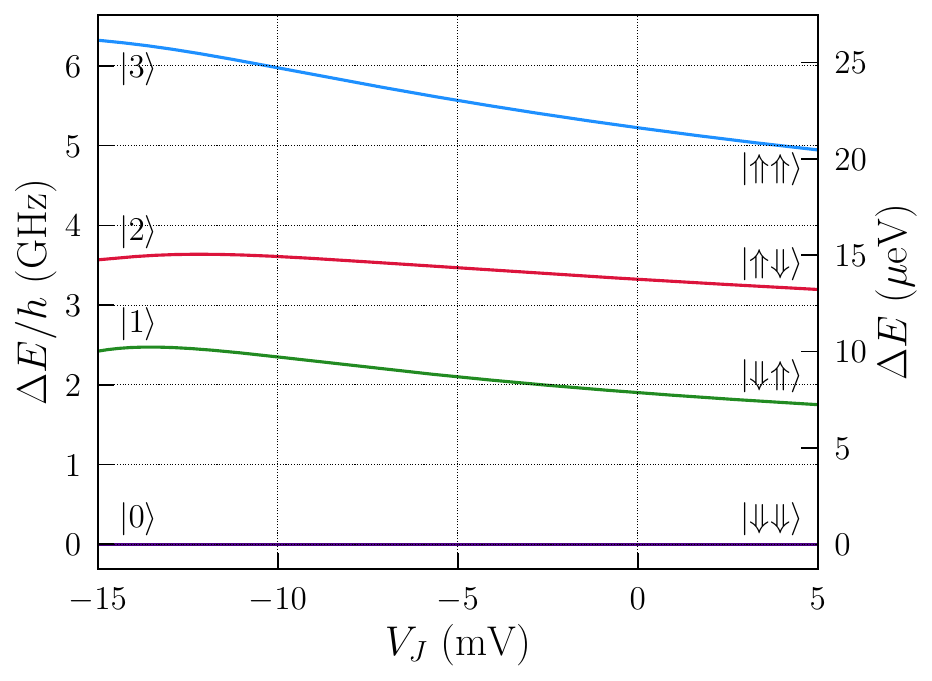}
	\caption{Splitting $\Delta E$ between the lowest two-hole levels and the ground-state as a function of the $J$-gate voltage $V_J$.}
	\label{fig:connection}
\end{figure}

According to Fig.~\ref{fig:connection}, $\ket{\Uparrow\Uparrow}$ is adiabatically connected by the $J$ gate to $\ket{3}$, $\ket{\Uparrow\Downarrow}$ to $\ket{2}$, $\ket{\Downarrow\Uparrow}$ to $\ket{1}$, and $\ket{\Downarrow\Downarrow}$ to $\ket{0}$. A slow enough $J$-gate ramp will thus map each computational basis state to the corresponding operational basis state, as required for best gate fidelity \cite{Russ18,Russ23}. However, a steep (diabatic) ramp will freeze the spin degrees of freedom (and ultimately the orbital ones), effectively projecting the computational onto the operational basis states. The intended one-to-one mapping between computational and operational basis states is then lost as the latter mix $\ket{\Uparrow\Downarrow}$ and $\ket{\Downarrow\Uparrow}$ components.

\begin{figure}[t]
	\centering
 	\includegraphics[width=\linewidth]{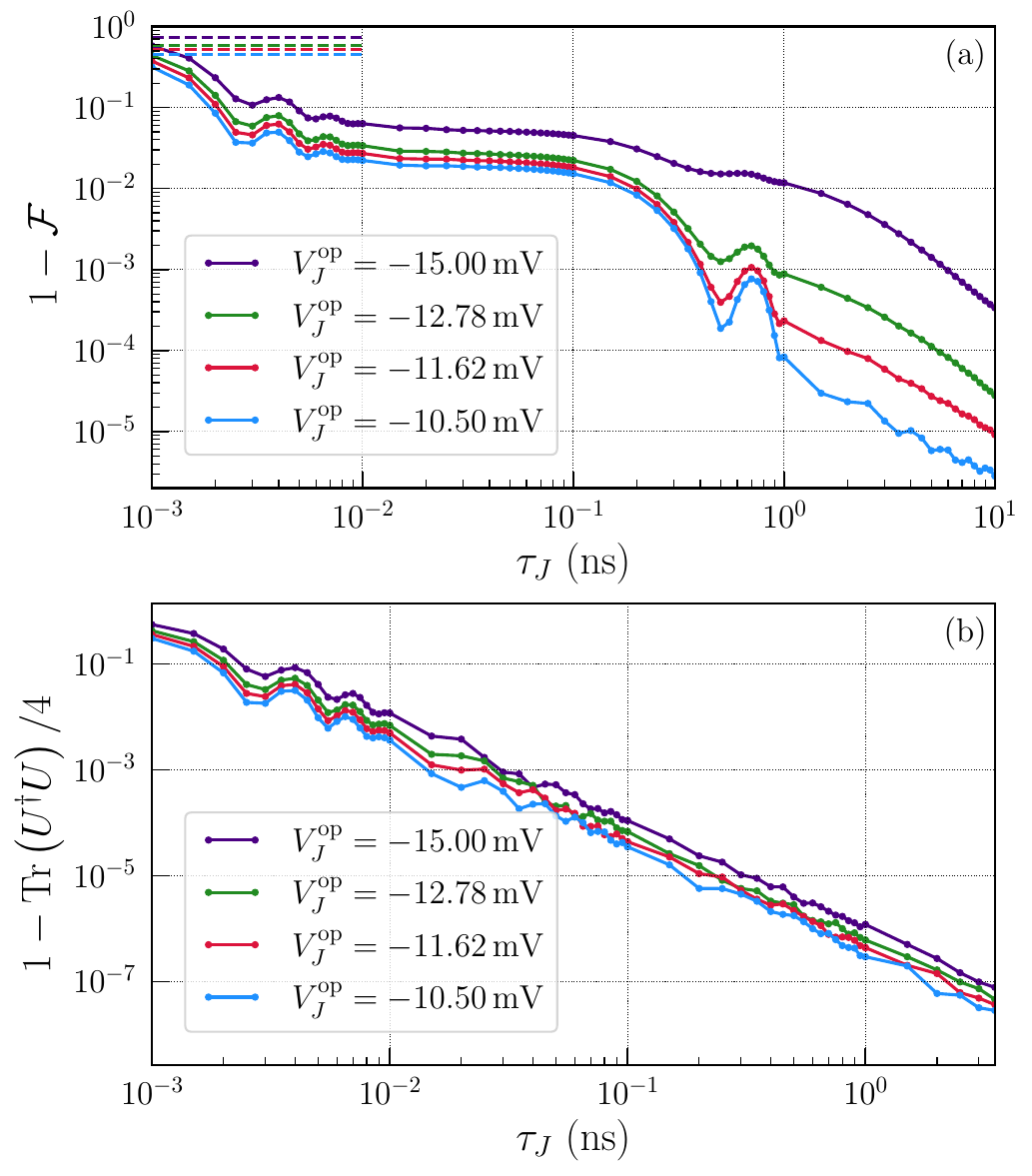}
	\caption{(a) Infidelity and (b) non-unitarity $1-\mathrm{Tr}(U^\dagger U)/4$ of a $J$-gate ramp $V_J=V_J^\mathrm{co}\to V_J^\mathrm{op}$ as a function of the ramp time $\tau_J$. The starting bias is $V_J^\mathrm{co}=5$\,mV (where the computational states are defined) and the final bias $V_J^\mathrm{op}$ (where the operational states are defined) is given in the legend (also see Table~\ref{tab:exchange} for the corresponding exchange energies). The horizontal dashed lines in (a) are the limiting fidelities when $\tau_J\to 0$.}
	\label{fig:Framp}
\end{figure}

The ideal ramp is thus expected to achieve the transformation
\begin{equation}
    U_{0,\mathrm{ramp}}=\begin{pmatrix}
		1 & 0 & 0 & 0 \\
		0 & e^{i\delta\chi_2} & 0 & 0 \\
		0 & 0 & e^{i\delta\chi_1} & 0 \\
		0 & 0 & 0 & e^{i\delta\chi_0}
	\end{pmatrix}
\end{equation}
between the input computational basis states (columns) and output operational basis states (rows). The $\delta\chi_i$ include an (irrelevant) contribution from the arbitrary phases chosen in the definition of the computational and operational basis states (but cancelled by the closing ramp), as well as a dynamical (and geometrical) contribution accumulated during the ramp (and not unravelled when closing the barrier). Assuming, as demonstrated in the previous section, that the latter can be corrected by phase gates, we define the fidelity of the ramp as:
\begin{equation}
    \mathcal{F}=\frac{1}{20}\left[\mathrm{Tr}\left(U^\dagger U\right)+\mathrm{Tr}\left(\abs{U}\right)^2\right]\,.
\end{equation}
The non-unitarity $1-\mathrm{Tr}(U^\dagger U)/4$ and the infidelity $1-\mathcal{F}$ of a ramp $V_J=V_J^\mathrm{co}\to V_J^\mathrm{op}$ are plotted as a function of the ramp time $\tau_J$ in Fig.~\ref{fig:Framp}. The ramp starts at $V_J^\mathrm{co}=5$\,mV (where the computational states are defined) and ends at different operational $V_J^\mathrm{op}$ (see Table~\ref{tab:exchange} for the corresponding exchange energies). We can distinguish three regimes: at very small $\tau_J<0.01$\,ns, both the infidelity and non-unitarity are large ($>0.1$). The ramp is, actually, diabatic with respect to both spin and charge degrees of freedom, so that the computational states leak out of the operational subspace. The horizontal dashed lines are the limiting fidelities ($\tau_J\to0$) computed from the projections of the computational onto the operational states (namely, assuming $\ket{\psi}\to\braket{0}{\psi}\ket{0}+\braket{1}{\psi}\ket{1}+\braket{2}{\psi}\ket{2}+\braket{3}{\psi}\ket{3}$ for any computational state $\ket{\psi}$). For intermediate $0.01\lesssim\tau_J\lesssim 0.1$\,ns, the ramp has become quasi-adiabatic with respect to charge; from there on, the non-unitarity decreases $\propto 1/\tau_J^2$, as expected from the lowest-order correction to the adiabatic theorem \cite{Kato50,Messiah61,Gu09}. Yet the infidelity reaches a plateau because the ramp is still diabatic with respect to spin. Finally, the infidelity drops distinctly once $\tau_J\gtrsim 1$\,ns, as the ramp becomes adiabatic with respect to spin and achieves the expected one-to-one correspondence between the computational and operational basis states.

We can estimate the transverse exchange energies from the plateau at $\tau_J\approx0.04$\,ns. We assume for that purpose that the computational states are mapped into the subspace of the operational states (adiabaticity with respect to charge), and are projected onto the eigenstates of $H_e$ [Eq.~\eqref{eq:Heff}] at the finite $J_\parallel$ and $J_\perp$ prevailing at $V_J=V_J^\mathrm{op}$ (diabaticity with respect to spin). The eigenenergies of this Hamiltonian are given by Eqs.~\eqref{eq:Es2}, and its eigenstates are the columns of the matrix:
\begin{equation}
    \renewcommand*{\arraystretch}{1.5}
	P=\frac{1}{\sqrt{2}}\begin{pmatrix}
		\sqrt{2} & 0 & 0 & 0\\
		0 & \frac{J_\perp}{\Delta\tilde{E}_Z}\frac{1}{\sqrt{1-\rho}} & \frac{-J_\perp}{\Delta\tilde{E}_Z}\frac{1}{\sqrt{1+\rho}} & 0\\
		0 & \sqrt{1-\rho} & \sqrt{1+\rho} & 0\\
		0 & 0 & 0 & \sqrt{2}
	\end{pmatrix}\,,
	\label{eq:PJperp}
\end{equation}
with $\Delta\tilde{E}_Z=\sqrt{\Delta E_Z^2+J_\perp^2}=E_2-E_1$ the splitting between the operational basis states $\ket{1}$ and $\ket{2}$, and $\rho=\Delta E_Z/\Delta\tilde{E}_Z$. We thus expect a transformation $U_\mathrm{ramp}$ of the form:
\begin{equation}
	U_\mathrm{ramp}\approx\begin{pmatrix}
		\approx 1 & 0 & 0 & 0\\
		0 & \alpha & \gamma & 0 \\
		0 & \beta & \delta & 0 \\
		0 & 0 & 0 & \approx 1
	\end{pmatrix}\equiv P^\dagger\,,
	\label{eq:Uramp}
\end{equation}
where $\abs{\alpha}^2+\abs{\beta}^2\approx 1$ and $\abs{\gamma}^2+\abs{\delta}^2\approx 1$. Matching Eqs.~\eqref{eq:PJperp} and \eqref{eq:Uramp} yields:
\begin{equation}
	J_\perp\approx J_\perp^*\equiv\Delta\tilde{E}_Z(|\alpha\gamma|+|\beta\delta|)\,.
	\label{eq:Jperp}
\end{equation}
The values of $J_\perp^*$ estimated from the calculated $U_\mathrm{ramp}$ matrices are reported in Table \ref{tab:exchange} for different operational $V_J^\mathrm{op}$, along with the corresponding $J_0$ and $J_\parallel$. They are close to the expected ${\bar J}_\perp=(J_0+J_\parallel)/2$. We emphasize, nevertheless, that Eq.~\eqref{eq:Uramp} misses significant mixings between the $\{\ket{\Uparrow\Uparrow},\ket{\Downarrow\Downarrow}\}$ and $\{\ket{\Uparrow\Downarrow},\ket{\Downarrow\Uparrow}\}$ subspaces captured by the numerical simulations. They result from couplings neglected in the derivation of the Hamiltonian $H_e$ [Eq.~\eqref{eq:Heff} and Appendix \ref{app:HundMulliken}], and from the dependence\footnote{This dependence is consistent with the leakage outside the computational subspace at small $\tau_J$, which shows that the computational and operational states do not only differ by their spin, but also by their orbital components.} of the $g$-factors (and Zeeman-split states $\ket{\Uparrow}$ and $\ket{\Downarrow}$) on $V_J$ (which is not accounted for in the above model) \cite{Martinez2026b}. This dependence is obvious in Fig.~\ref{fig:connection}, where, for example, the transition energy $E_3-E_0=\mu_bB(g_L^*+g_R^*)$ increases with negative $V_J$.

As expected, the larger $J_\perp$ (which mixes $\ket{\Uparrow\Downarrow}$ and $\ket{\Downarrow\Uparrow}$), the smaller the fidelity of the ramp for a given $\tau_J$. The ramp must thus be slower for larger transverse exchange energies $J_\perp$. As an illustration, the ramp time to achieve an infidelity better than $10^{-3}$ must be greater than $6$\,ns for $V_J^\mathrm{op}=-15$\,mV, $0.9$\,ns for $V_J^\mathrm{op}=-12.78$\,mV, $0.45$\,ns for $V_J^\mathrm{op}=-11.62$\,mV and $0.40$\,ns for $V_J^\mathrm{op}=-10.50$\,mV. However, the smaller longitudinal exchange energies $J_\parallel$ that usually come with weaker $J_\perp$ may compromise the fidelity of the $\pi$-pulse as discussed below. 

We analyze the fidelity of the phase gate ramps in Appendix \ref{app:phasegateramps}. Given the small span of these ramps ($1$\,mV), they are quasi-adiabatic once $\tau_r\gtrsim 0.5$\,ns, and the infidelity hardly exceeds $10^{-4}$ down to picosecond ramps.

\subsection{Fidelity of the $\pi$-pulse}
\label{sec:drive}

In this section, we discuss the fidelity of the $\pi$-pulse, and the management of off-resonant couplings.

First, we remind that the probability of transition between an initial state $\ket{-}$ (with energy $E_-$) and a state $\ket{+}$ (with energy $E_+$) under a RF drive with amplitude $V_\mathrm{ac}$ and frequency $f$ is \cite{Messiah61}
\begin{equation}
	\mathrm{P}_+(t)=P^\mathrm{max}\sin^2\left(\pi{\tilde\nu}t\right),
\end{equation}
where 
\begin{subequations}
\label{eq:Rabi}
\begin{align}
    {\tilde\nu}&=\sqrt{\nu^2+\Delta f^2} \\
    P^\mathrm{max}&=\left(\frac{\nu}{\tilde\nu}\right)^2
\end{align}
\end{subequations}
are respectively the Rabi frequency and the amplitude of the Rabi oscillations, with $\Delta f=f-f_0$ the detuning of the drive with respect to the resonance frequency $f_0=(E_+-E_-)/h$. In the linear response regime, the resonant Rabi frequency is $\nu=\abs{\bra{+}D\ket{-}}/h=\lambda V_\mathrm{ac}/h$, where $D\propto V_\mathrm{ac}$ is the operator describing the drive in the Hamiltonian. The amplitude $P^\mathrm{max}$ of the Rabi oscillations is thus unity at resonance and decreases as a Lorentzian off-resonance while their frequency $\tilde\nu$ increases.

\subsubsection{Choice of the driving gate}

The choice of the driving gate is critical for the operation of the CNOT gate. First, as discussed in Ref. \cite{martinez2022hole,Abadillo2023,Valentin25}, the right plunger gate $P_R$ is inefficient in driving the right dot spin for in-plane magnetic fields. Indeed, the quasi-circular $R$ dot wave function primarily breathes in the electric field of that gate, which modulates the effective in-plane $g$-factors $g_x^*\approx g_y^*$ (longitudinal coupling) but hardly gives rise to Rabi oscillations (transverse coupling). For a magnetic field $\vec{B}\parallel\vec{x}$, the Rabi frequency is expected to be maximal when the drive field is also parallel to $\vec{x}$, as is essentially the case for a modulation of the left plunger ($V_L$) or right barrier ($V_{RR}$) gate voltages.

\begin{figure}[t]
	\centering
   	\includegraphics[width=\linewidth]{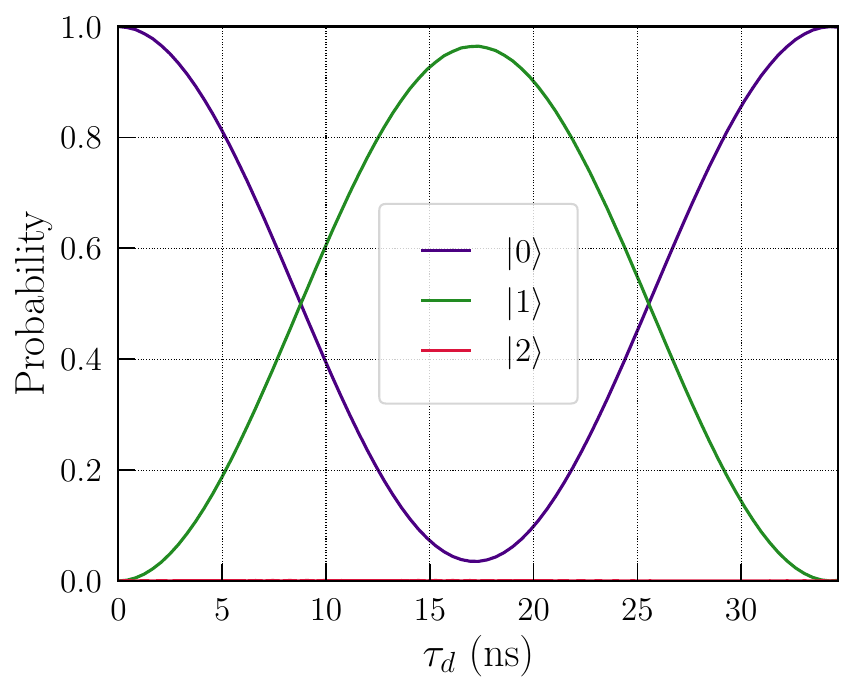}
	\caption{Probability of being in the $\ket{0}$, $\ket{1}$ and $\ket{2}$ states as a function of the drive time $\tau_d$. The DQD is driven with a RF pulse of amplitude $V_\mathrm{ac}=1$\,mV and frequency $f=f_\Downarrow=(E_1-E_0)/h$ on the left plunger gate $P_L$ ($V_J^\mathrm{op}=-15$\,mV). The initial state at $\tau_d=0$ is $\ket{0}$.}
	\label{fig:driveL}
\end{figure}

We can thus attempt to drive the $\ket{0}\leftrightarrow\ket{1}$ transition with a RF pulse with amplitude $V_\mathrm{ac}=1$\,mV and frequency $f=f_\Downarrow=(E_1-E_0)/h$ on the left plunger gate $P_L$. We plot in Fig.~\ref{fig:driveL} the probability of being in the $\ket{0}$, $\ket{1}$ and $\ket{2}$ states as a function of the drive time $\tau_d$ (the initial state at $\tau_d=0$ being $\ket{0}$). The Rabi oscillations between $\ket{0}$ and $\ket{1}$ are prominent, with little leakage to the nearby $\ket{2}$ state. Yet these oscillations are not fully visible (the system never reaches the $\ket{1}$ state). This results from the AC Stark shift \cite{Barrat61,Bakos77} of the $\ket{0}\leftrightarrow\ket{1}$ transition by off-resonant coupling to excited states (mostly) outside the operational subspace. 

Let $\ket{m}$ and $E_m$ be the eigenstates and eigenenergies at the operational bias point ($0\le m\le3$ being the operational basis states). The off-resonant couplings\footnote{This expression includes both co-rotating and counter-rotating contributions; for completeness, we may add the counter-rotating shift of the driven transition (the Bloch-Siegert shift \cite{Bloch40}) $\delta f_\Downarrow^\prime=\abs{D_{10}}^2/(4h\Delta_{10})$, which is however very small.} $D_{mn}=\bra{m}D\ket{n}$ between the $n=0,1$ and the $m>1$ states indeed shift the resonance frequency $f_\Downarrow=E_1-E_0$ by \cite{Bakos77}
\begin{align}
    \delta f_\Downarrow=\frac{1}{2h}\sum_{m>1}&\left(\abs{D_{m0}}^2\frac{\Delta_{m0}}{\Delta_{m0}^2-\Delta_{10}^2}-\right. \nonumber \\ &\left.\abs{D_{m1}}^2\frac{\Delta_{m1}}{\Delta_{m1}^2-\Delta_{10}^2}\right)\propto V_\mathrm{ac}^2\,.
    \label{eq:ACstark}
\end{align}
where $\Delta_{mn}=E_m-E_n$. If not compensated by a careful tuning of the drive frequency, this AC Stark shift can result in a significant decrease $\delta P^\mathrm{max}$ of the amplitude of the Rabi oscillations:
\begin{equation}
    \delta P^\mathrm{max}=P^\mathrm{max}-1=-\frac{\delta f_\Downarrow^2}{\delta f_\Downarrow^2+\nu^2}\,.
\end{equation}
This is particularly sensitive for slow Rabi oscillations with narrow linewidth ($\nu=\abs{D_{01}}/h\lesssim\abs{\delta f_\Downarrow}$). From the time traces of Fig.~\ref{fig:driveL}, we estimate $\tilde\nu=28.83$\,MHz and $P^\mathrm{max}=0.965$, thus $\abs{\delta f_\Downarrow}={\tilde\nu}\sqrt{1-P^\mathrm{max}}=5.39$\,MHz. This is consistent with Eq.~\eqref{eq:ACstark}, which yields $\delta f_\Downarrow=-5.52$\,MHz (of which $93$\% comes from $m>3$ excitations outside the operational subspace). We have checked that driving the $P_L$ gate at frequency $f=f_\Downarrow-5.5$\,MHz indeed improves\footnote{We emphasize that off-resonant couplings can still limit the amplitude of the Rabi oscillations due to coherent, $\propto(\abs{D_{m0}}/\Delta_{m0})^2$ leakages, as explained in section \ref{sec:offresonant}.} the visibility of the Rabi oscillations ($P^\mathrm{max}\approx 1$).

The AC Stark shift scales as $V_\mathrm{ac}^2$ and may thus complicate the calibration of the CNOT gate (see next section about the optimization of $V_\mathrm{ac}$). It depends on the driving gate through the pattern of off-resonant couplings $\abs{D_{m0}}$ and $\abs{D_{m1}}$ with the excited states. We can therefore compare the fidelities of different driving gates (without drive frequency correction) to identify which one is more robust to AC Stark shifts. As for the ramps, we do not care for the phases accumulated during the drive, and thus define the $\pi$-pulse fidelity as:
\begin{align}
    \mathcal{F}&=\frac{1}{20}\left[\mathrm{Tr}\left(U^\dagger U\right)+\abs{\mathrm{Tr}\left(U_0^\dagger\abs{U}\right)}^2\right] \nonumber \\
    &=\frac{1}{20}\left[\mathrm{Tr}\left(U^\dagger U\right)+(\abs{U_{11}}+\abs{U_{22}}+\abs{U_{34}}+\abs{U_{43}})^2\right]\,,
    \label{eq:Fdrive}
\end{align}
with $U_0$ given by Eq.~\eqref{eq:U0}. The matrix $U$ is calculated in the operational subspace (the system is initialized in the $\{\ket{0},\ket{1},\ket{2},\ket{3}\}$ states, and projected on the same states after the drive). 

\begin{figure}[t]
	\centering
  	\includegraphics[width=\linewidth]{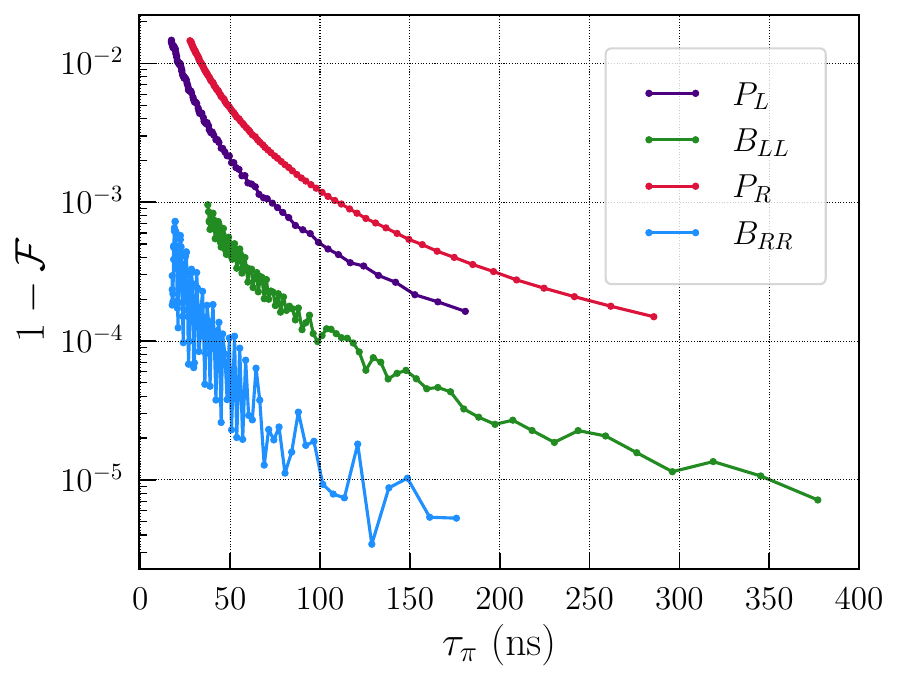}
	\caption{Infidelity $1-\mathcal{F}$ of the $\pi$-pulse as a function of the drive time $\tau_\pi\propto 1/V_\mathrm{ac}$, for drives with the gates $P_L$, $P_R$, $B_{RR}$ and $B_{LL}$ ($V_J^\mathrm{op}=-15$\,mV). The drive frequency $f=f_\Downarrow=(E_1-E_0)/h$ is not corrected for the AC Stark shift.}
	\label{fig:Fdrivegates}
\end{figure}

The fidelity of the $\pi$-pulse is plotted as a function of $\tau_\pi\propto 1/V_\mathrm{ac}$ in Fig.~\ref{fig:Fdrivegates}, for drives with the gates $P_L$, $P_R$, $B_{RR}$ and $B_{LL}$ ($V_J^\mathrm{op}=-15$\,mV). The infidelity tends to increase the faster the smaller $\tau_\pi$ as off-resonant processes depend non-linearly on the drive voltage $V_\mathrm{ac}$. There is thus a trade-off between the $\pi$-pulse fidelity $\mathcal{F}$ and duration $\tau_\pi$.

The different gates do not, as expected, perform similarly. The right plunger gate $P_R$ exhibits the worst infidelities $1-\mathcal{F}\gtrsim 1\%$ when $\tau_\pi<30$\,ns. As discussed above, the coupling $\abs{D_{01}}$ of the intended $\ket{0}\leftrightarrow\ket{1}$ transition is small --- much smaller than the couplings $\abs{D_{m0}}$ and $\abs{D_{m1}}$ with far detuned excitations. As a consequence, $V_\mathrm{ac}$ must be large to reach a reasonably short $\tau_\pi$, which enhances the AC Stark shift. These ratios improve for the left plunger gate $P_L$, and even for the remote $B_{LL}$ gate. Nevertheless, the best driving gate in the present layout is $B_{RR}$, which is the most strongly coupled to the $\ket{0}\leftrightarrow\ket{1}$ transition in the right dot and the least coupled to the left dot. It also shows much smaller AC Stark shifts ($\delta f_\Downarrow<1$\,MHz for $V_\mathrm{ac}<1$\,mV). This allows for fast $\pi$-pulses at small $V_\mathrm{ac}$ where non-linearities remain negligible.

The infidelity of the $B_{RR}$ gate is, however, oscillating with $\tau_\pi/V_\mathrm{ac}$. We discuss the cause (and cure) for these oscillations in the next section.

\subsubsection{Off-resonant drive of the $\ket{2}\leftrightarrow\ket{3}$ transition}
\label{sec:offresonant}

One of the main source of residual infidelity is the unwanted drive of the closest $\ket{2}\leftrightarrow\ket{3}$ transition, detuned from the intended $\ket{0}\leftrightarrow\ket{1}$ transition by $\Delta f=f_\Uparrow-f_\Downarrow=[(E_3-E_2)-(E_1-E_0)]/h=J_\parallel/h$. This results in a rotation of the spin of the $R$ dot even though the spin of the $L$ dot is up \cite{Russ18}. 

\begin{figure*}[t]
    \centering
    \includegraphics[width=0.8\linewidth]{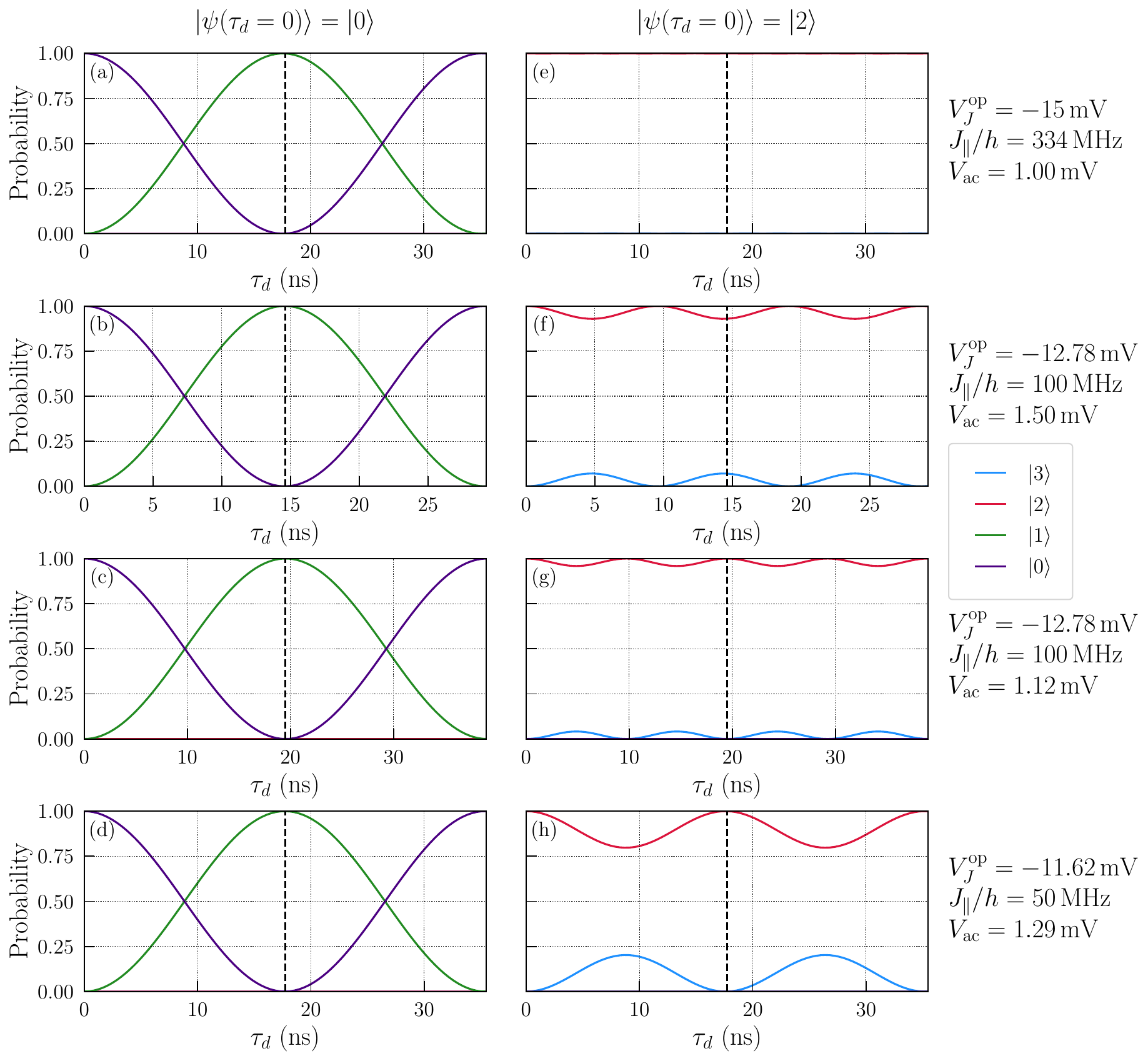}
    \caption{Probability of being in the $\{\ket{0},\ket{1},\ket{2},\ket{3}\}$ states as a function of the drive time $\tau_d$, starting from either (a)-(d) $\ket{0}$ or (e)-(h) $\ket{2}$ at $\tau_d=0$, for different operational $V_J^\mathrm{op}$ (different $J_\parallel$) and drive amplitudes $V_\mathrm{ac}$ on the $B_{RR}$ gate. The vertical dashed lines indicate the $\pi$-pulse time $\tau_\pi$.}
    \label{fig:Rabi23}
\end{figure*}

This dynamics is illustrated by Fig.~\ref{fig:Rabi23}, which plots the probability of being in the $\{\ket{0},\ket{1},\ket{2},\ket{3}\}$ states as a function of the drive time $\tau_d$, starting from either $\ket{0}$ or $\ket{2}$ at $\tau_d=0$, for different operational $V_J^\mathrm{op}$ (different $J_\parallel$) and $V_\mathrm{ac}$. The main, resonant Rabi oscillations between $\ket{0}$ and $\ket{1}$ are well visible on the left panels (a)-(d). Their amplitude $P_{01}^\mathrm{max}$ is $\approx 1$ as required for a high-fidelity CNOT. The target $\pi$-pulse is thus completed in $\tau_\pi=1/(2\nu_{01})$ with $\nu_{01}=\lambda_{01}V_\mathrm{ac}/h$ the Rabi frequency. However, undesirable off-resonant oscillations between $\ket{2}$ and $\ket{3}$ can also be spotted on panels (e)-(h). According to Eqs.~\eqref{eq:Rabi}, the amplitude and frequency of these oscillations are:
\begin{subequations}
\begin{align}
    P_{23}^\mathrm{max}&=\frac{\nu_{23}^2}{\nu_{23}^2+(J_\parallel/h)^2} \\
    {\tilde\nu}_{23}&=\sqrt{\nu_{23}^2+(J_\parallel/h)^2}\,
\end{align}
\end{subequations}
with $\nu_{23}=\lambda_{23}V_\mathrm{ac}/h$ the resonant Rabi frequency of the $\ket{2}\leftrightarrow\ket{3}$ oscillations. They are barely noticeable at large $J_\parallel$ but become more prominent as $J_\parallel$ decreases or $V_\mathrm{ac}$ increases. Their amplitude indeed scales as $P_{23}^\mathrm{max}\approx (h\nu_{23}/J_\parallel)^2=(\lambda_{23}V_\mathrm{ac}/J_\parallel)^2$.

Such off-resonant oscillations can only be suppressed by reducing $V_\mathrm{ac}$, at the cost of increasing the gate time $\tau_\pi$. Alternatively, $V_\mathrm{ac}$ can be chosen so that the $\ket{2}\leftrightarrow\ket{3}$ oscillations complete an integer number of periods within $\tau_\pi$ (synchronization with the $\ket{0}\leftrightarrow\ket{1}$ oscillations) \cite{Russ18,Heinz22,Russ23}. That way, $\ket{2}$ and $\ket{3}$ are left invariant (up to a phase) by the drive. This leads to the condition
\begin{equation}
    {\tilde\nu}_{23}\tau_\pi=\sqrt{\nu_{23}^2+(J_\parallel/h)^2}\tau_\pi=n
\end{equation}
with $n$ an integer. Substituting $\nu_{23}=\lambda_{23}V_\mathrm{ac}/h$ and $\tau_\pi=1/(2\nu_{01})=h/(2\lambda_{01}V_\mathrm{ac})$ yields:
\begin{equation}
    V_\mathrm{ac}=\frac{\abs{J_\parallel}}{\sqrt{\left(2n\lambda_{01}\right)^2-\lambda_{23}^2}}\,.
	\label{eq:Vacs}
\end{equation}
This condition is illustrated by panels (g) and (h) of Fig.~\ref{fig:Rabi23} [$V_\mathrm{ac}=1.1$\,mV for $n=2$ at $V_J^\mathrm{op}=-12.78$\,mV ($J_\parallel/h=100$\,MHz), and $V_\mathrm{ac}=1.27$\,mV for $n=1$ at $V_J^\mathrm{op}=-11.62$\,mV ($J_\parallel/h=50$\,MHz)]. The chosen values of $n$ are the smallest such that $V_\mathrm{ac}\lesssim 1.5$\,mV to limit non-linearities. 

\begin{figure}[t]
    \centering
    \includegraphics[width=\linewidth]{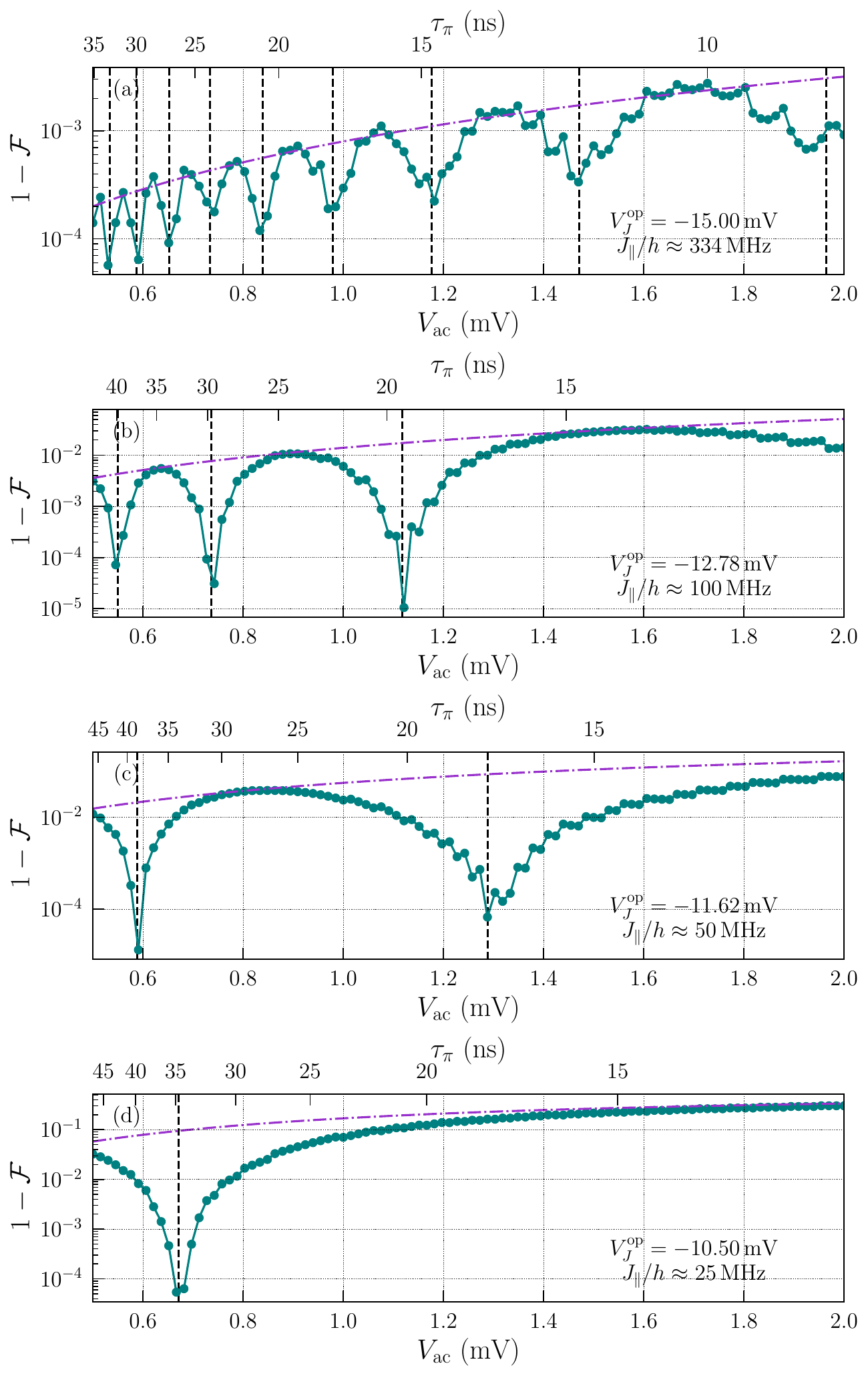}
    \caption{Infidelity of the $\pi$-pulse as a function of the drive amplitude $V_\mathrm{ac}$ on the $B_{RR}$ gate for different operational points. The corresponding $\tau_\pi$ is reported as the secondary axis on top. The vertical dashed lines are the condition \eqref{eq:Vacs}. The purple dash-dotted line is the envelope of the infidelity $1-{\cal F}=[4-(1+\sqrt{1-P_{23}^\mathrm{max}})^2]/5$.}
    \label{fig:FVac}
\end{figure}

We finally plot in Fig.~\ref{fig:FVac} the infidelity\footnote{The small steps at short $\tau_\pi$ in Fig.~\ref{fig:FVac} result from the rounding of $\tau_\pi$ to the nearest number of Larmor periods $T_\Downarrow=1/f_\Downarrow$ in order to avoid any discontinuity in the drive signal.} of the $\pi$-pulse as a function of the drive amplitude $V_\mathrm{ac}$ on the $B_{RR}$ gate, for different values of $J_\parallel$. The corresponding $\tau_\pi$ is reported as a secondary axis on top. The trends discussed above are well highlighted by this figure. First of all, the overall envelope of the infidelity is\footnote{In the worst case scenario, the probability of being in $\ket{3}$ starting from $\ket{2}$ is $P_{23}^\mathrm{max}$ at the end of the drive. Assuming that the $\ket{0}\leftrightarrow\ket{1}$ transition is perfectly driven, the matrix of the operation at $\tau_d=\tau_\pi$ is then:
\begin{equation}
    \abs{U}=\begin{pmatrix}
        \sqrt{1-P_{23}^\mathrm{max}} & \sqrt{P_{23}^\mathrm{max}} & 0 & 0\\
        \sqrt{P_{23}^\mathrm{max}} & \sqrt{1-P_{23}^\mathrm{max}} & 0 & 0\\
        0 & 0 & 0 & 1\\
        0 & 0 & 1 & 0
    \end{pmatrix}.
\end{equation}
Following Eq.~\eqref{eq:Fdrive}, we reach:
\begin{equation}
    1-\mathcal{F}=\frac{4-\left(1+\sqrt{1-P_{23}^\mathrm{max}}\right)^2}{5}\,.
\end{equation}} $1-{\cal F}=[4-(1+\sqrt{1-P_{23}^\mathrm{max}})^2]/5$ (the purple dash-dotted line). It however shows distinct dips each time the condition \eqref{eq:Vacs} is satisfied (dashed vertical lines). These dips get thinner as their index $n$ increases (thus $V_\mathrm{ac}$ decreases) because the $\ket{0}\leftrightarrow\ket{1}$ oscillations get slower and slower relative to the unwanted $\ket{2}\leftrightarrow\ket{3}$ oscillations. This figure demonstrates that high-fidelity CNOT gates can be achieved even at small $J_\parallel$, at the cost of an accurate calibration of the drive amplitude $V_\mathrm{ac}$. As discussed in section \ref{sec:Jramps}, a smaller $J_\perp$ also favors the adiabaticity of the $J$-gate ramps, allowing for faster opening/closing of the barrier. We emphasize, nevertheless, that for given $\lambda_{01}$ and $\lambda_{23}$ the optimal $V_\mathrm{ac}$ are $\propto J_\parallel$ for a given $n$. Therefore, the largest optimal $V_\mathrm{ac}$ [smallest $n\ge1$ such that $(2n\lambda_{01})^2-\lambda_{23}^2>0$] may remain small at low $J_\parallel$, making the total gate time longer. 

\subsection{Discussion}

We further discuss the fidelity of the combined $J$-gate ramps and $\pi$-pulse. We show in Appendix \ref{app:timetraces} some time traces highlighting the effect of the diabaticity of the ramps on the drive.

\begin{figure}[t]
	\centering
	\includegraphics[width=\linewidth]{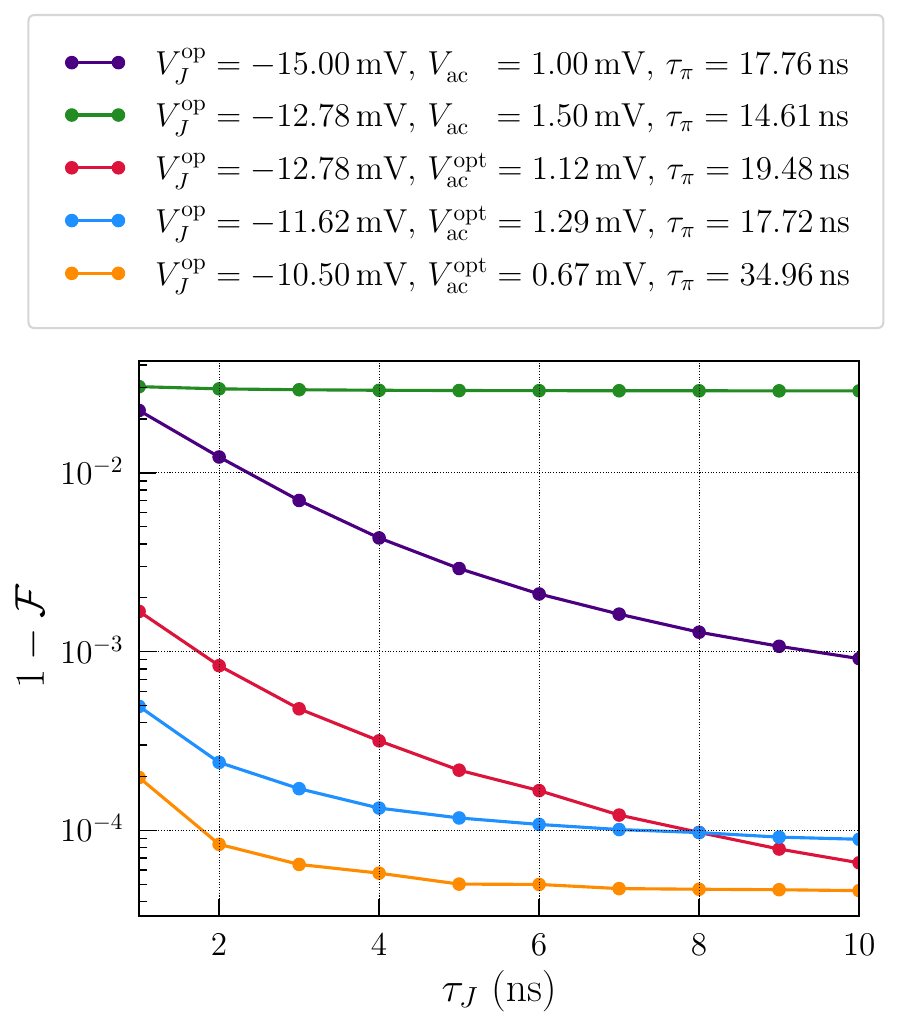}
	\caption{Infidelity of the combined $J$-gate ramps and $\pi$ pulse as a function of the ramp time $\tau_J$, for different operational bias $V_J^\mathrm{op}$ and driving amplitudes $V_\mathrm{ac}$. The driving amplitudes $V_\mathrm{ac}^\mathrm{opt}$ have been optimized to reduce the impact of off-resonant transitions (section \ref{sec:offresonant}).}
	\label{fig:InfidFull}
\end{figure}

We plot the infidelity of the CNOT gate in Fig.~\ref{fig:InfidFull} as a function of the $J$-gate ramp time $\tau_J$ for different values of $V_J^\mathrm{op}$ and $V_\mathrm{ac}$. We still assume that the phase drifts can be corrected exactly by the phase gates and use Eq.~\eqref{eq:Fdrive} for the fidelity in the computational subspace (the system is initialized in a computational state, then the barrier is opened, the $B_{RR}$ gate is driven, the barrier is closed and the final state is projected into the computational subspace). The degradation of the fidelity with decreasing $\tau_J$ is due to the increasing diabaticity of the $J$-gate ramps discussed in section \ref{sec:Jramps} (the computational states do not map one-to-one onto the operational states when the barrier is opened/closed). At large $J_\parallel=334$\,MHz ($V_J^\mathrm{op}=-15$\,mV), the infidelity steadily decreases with increasing $\tau_J$ (down to the $10^{-3}$ range) whatever $V_\mathrm{ac}<1.25$\,mV. However, at smaller $J_\parallel=100$\,MHz ($V_J^\mathrm{op}=-12.78$\,mV), the infidelity quickly saturates for $V_\mathrm{ac}=1.5$\,mV, as the error gets dominated by the spurious $\ket{2}\leftrightarrow\ket{3}$ transitions (see Fig.~\ref{fig:FVac}b). Nevertheless, these transitions can be suppressed by a suitable choice of drive amplitude $V_\mathrm{ac}=1.12$\,mV, as shown in the previous section. At such an optimal $V_\mathrm{ac}$, the infidelity drops considerably over the whole range of $\tau_J$, and can be lower than $10^{-4}$ when $\tau_J>8$\,ns. Better fidelities can even be achieved at short $\tau_J$ for smaller $J_\parallel=25$--$50$\,MHz (but optimal $V_\mathrm{ac}$) because $J_\perp$ is also weaker (Table \ref{tab:exchange}), which reduces the diabaticity of the ramps. We emphasize, though, that infidelities in the $<10^{-4}$ range may be limited by various calibration errors and may thus not be physically meaningful.

In conclusion, the fidelity of the CNOT gate is limited by the interplay between the diabaticity of the mapping between the computational and operational bases, and by the emergence of off-resonant couplings and Rabi oscillations. The former is mitigated by increasing the $J$-gate ramp times and/or decreasing the exchange energies, while the latter call for a suitable choice of driving gate and for a careful optimization of the drive amplitude $V_\mathrm{ac}$ if the exchange energy is small. In principle (and in the absence of noise), faithful CNOT gates can be achieved down to small $J_\parallel<30$\,MHz. However, the $\pi$-pulse time $\tau_\pi$ may become very long as the largest optimal $V_\mathrm{ac}$ scales with $J_\parallel$, which ultimately cancels the gains on the ramp time $\tau_J$. As an illustration, the fastest, optimized drive time increases from $\tau_\pi=17.72$\,ns to $\tau_\pi=34.96$\,ns when $J_\parallel$ goes from $50$\,MHz to $25$\,MHz, without much benefit for the adiabaticity of the $J$-gate ramps (see Fig.~\ref{fig:InfidFull}). The range $J_\parallel\approx 50$--$100$\,MHz does, therefore, provide the best balance between ramp diabaticity, drive errors and gate times in this device. For $J_\parallel=100$\,MHz for example ($V_J^\mathrm{op}=-12.78$\,mV), $\mathcal{F}=99.95$\% (including the phase gates) and $\tau_\mathrm{CNOT}=61$\,ns at $V_\mathrm{ac}=1.12$\,mV, enabling fast and accurate CNOT gates ($\tau_J=5$\,ns, $\tau_\pi=19.5$\,ns, $\tau_w=6.5$\,ns, $\tau_r=1$\,ns, $\tau_1=5$\,ns and $\tau_2=17$\,ns).

\section{The CNOT in the noisy device}
\label{sec:noisyCNOT}

In this section, we discuss the effects of quasi-static charge noise on the fidelity of the CNOT gate. We consider two kinds of charge fluctuations: {\it i}) A neutral trap at the GeSi/Al$_2$O$_3$ interface capturing/releasing an electron from/to a gate and {\it ii}) a charge $q$ moving over some distance $d$ at the GeSi/Al$_2$O$_3$ interface \cite{Martinez2022,Martinez24,Martinez2026,Shehata23}. The first case can be modeled as a point charge at the interface, and the second as a dipole $p=qd$.

We can compute the potential of such point charges (assumed positive here) and dipoles with the Poisson solver\footnote{These potentials of the charges and dipoles are computed with all gates grounded.} and add it to the time-dependent Hamiltonian. In a first approximation, a charge trapped below a gate actually makes a dipole with its image charge in that gate, and a dipole under a gate likewise forms a quadrupole with its image charges. We consider the limit of quasi-static noise, where the dynamics of the traps is very slow compared to the duration of the CNOT gate. Therefore, we will not deal with the (rare) events where a trap switches during the execution of the CNOT gate, but focus on the effects of such a switch on all subsequent CNOT operations. We thus assess the fidelity of the CNOT gate in the presence of a point charge or dipole at the GeSi/Al$_2$O$_3$ interface, with the gate parameters ($V_\mathrm{ac}$, $\tau_J$, $\tau_\pi$, $\tau_w$, $\tau_r$, $\tau_1$, $\tau_2$) calibrated in the pristine device.

The fidelity needs to be calculated in the same computational subspace in the pristine and disordered devices. We compute, therefore, the evolution of the computational states $\{\ket{\Uparrow\Uparrow},\ket{\Uparrow\Downarrow},\ket{\Downarrow\Uparrow},\ket{\Downarrow\Downarrow}\}$ of the pristine device in the presence of the charge or dipole, and project the outcome back onto these computational states to build the transformation matrix $U$. We define two fidelities:
\begin{subequations}
\begin{align}
    \mathcal{F}_0&=\frac{1}{20}\left[\mathrm{Tr}\left(U^\dagger U\right)+\abs{\mathrm{Tr}\left(U_0^\dagger U\right)}^2\right] \\
    \mathcal{F}^*&=\frac{1}{20}\left[\mathrm{Tr}\left(M^\dagger M\right)+\abs{\mathrm{Tr}\left(M\right)}^2\right]\,,
\end{align}
\end{subequations}
where $U_0$ is the transformation matrix of an ideal CNOT [Eq.~\eqref{eq:U0}], while $M$ is the matrix of the projection of the final states in the disordered device (columns) onto the final states in the pristine device (lines). $\mathcal{F}_0$ thus measures how close the operation in the disordered device is to an ideal CNOT gate, while $\mathcal{F}^*$ measures how close it is to the operation in the pristine device ($\mathcal{F}^*=1$ if they are identical). The fact that the computational subspaces of the pristine and disordered devices are practically different is seen as leakage.

\begin{figure}[t]
    \centering
 	\includegraphics[width=\linewidth]{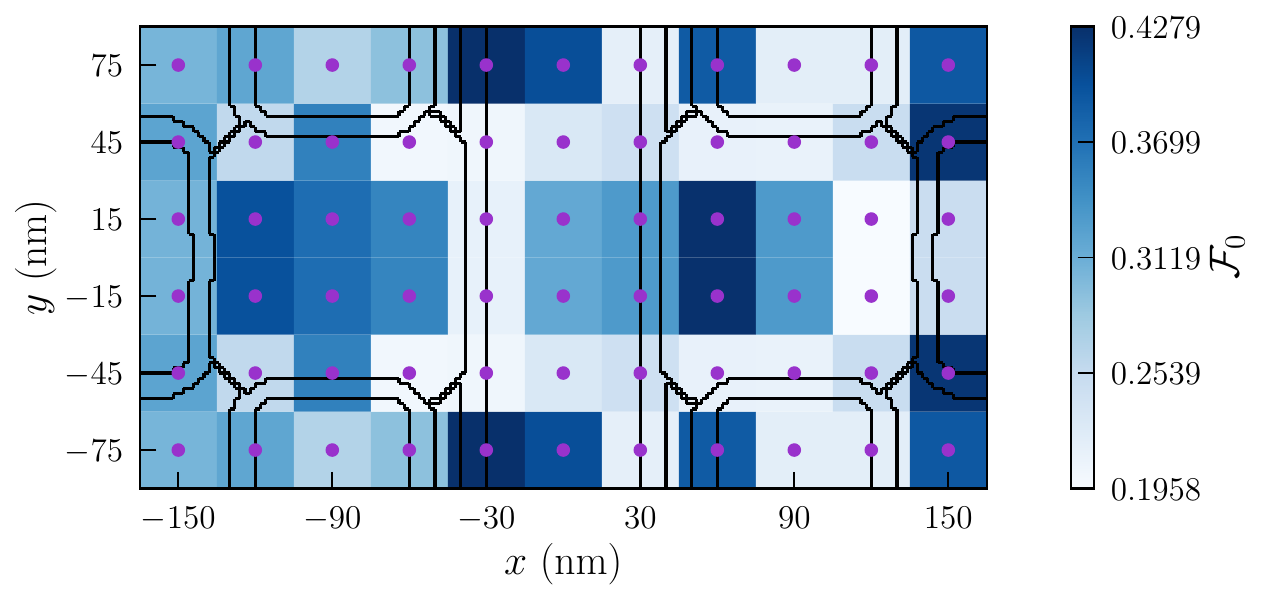}
    \caption{Fidelities $\mathcal{F}_0$ of the CNOT gate in the presence of a charge $+e$ trapped at the GeSi/Al$_2$O$_3$ interface as a function of its position in the $(xy)$ plane (marked by the purple dots). The operational $J$-gate voltage is $V_J^\mathrm{op}=-12.78$\,mV where $J_\parallel=100$\,MHz (optimized $V_\mathrm{ac}=1.12$\,mV and $\tau_J=5$\,ns, $\tau_\pi=19.5$\,ns, $\tau_w=6.5$\,ns, $\tau_r=1$\,ns, $\tau_1=5$\,ns, $\tau_2=17$\,ns in the pristine device). The position of the gates is delineated by the dark lines.}
    \label{fig:charge}
\end{figure}

\begin{figure}[t]
    \centering
 	\includegraphics[width=\linewidth]{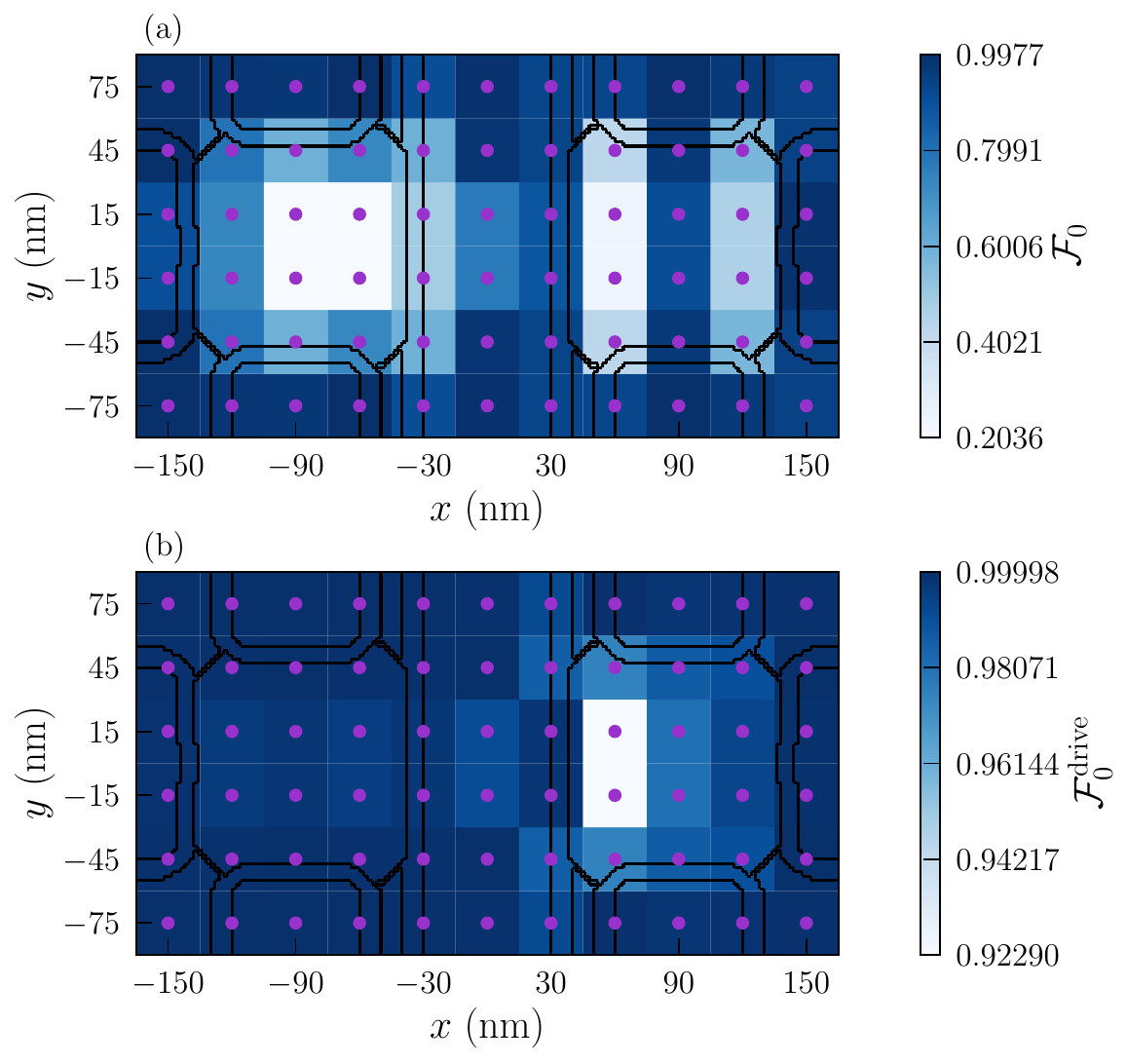}
    \caption{Fidelities (a) $\mathcal{F}_0$ and (b) $\mathcal{F}_0^\mathrm{drive}$ of the CNOT gate in the presence of a dipole $\vec{p}=(1\,e\text{.nm})\vec{x}$ at the GeSi/Al$_2$O$_3$ interface as a function of its position in the $(xy)$ plane (marked by the purple dots). The bias point and timings are the same as in Fig.~\ref{fig:charge}. The position of the gates is delineated by the dark lines.}
    \label{fig:dipole}
\end{figure}

We plot $\mathcal{F}_0$ as a function of the position of the charge or dipole at the interface in Figs.~\ref{fig:charge} and \ref{fig:dipole}a. The dipoles $p=1\,e$.nm are oriented along $x$. The operational $J$-gate voltage is $V_J^\mathrm{op}=-12.78$\,mV ($J_\parallel=100$\,MHz), and the fidelity of the pristine device (including the phase gates) is $\mathcal{F}_0=99.95$\% (for a total gate time $\tau_\mathrm{CNOT}=61$\,ns). 

The fidelity of the CNOT gate is considerably degraded by the capture of a charge in the vicinity of the DQD. It can be as small\footnote{Actually, $\mathcal{F}_0=0.2$ is the minimum fidelity of a unitary process, where $\mathrm{Tr}\left(U^\dagger U\right)=1$ but $\abs{\mathrm{Tr}\left(U_0^\dagger U\right)}^2=0$. However, the transformation is not even unitary in the presence of charge traps; $1-\mathrm{Tr}\left(U^\dagger U\right)/4$ is indeed up to 7\% on the map of Fig.~\ref{fig:charge} due to the mismatch between the computational subspaces with and without the charge. Non-unitarity is, nevertheless, negligible for dipoles on the map of Fig.~\ref{fig:dipole} [$1-\mathrm{Tr}\left(U^\dagger U\right)/4<10^{-4}$].}  as $\approx 0.2$ (in the investigated area) but the map does not show any clear pattern beyond the $(xz)$ mirror expected from the symmetry of the structure. This results from sizable shifts of the energy levels, which detune the drive and dephase the qubits. We conclude, therefore, that charge trapping events occurring below the gates {\it after} the CNOT calibration are prohibitive.

The situation is more favorable for (moderate) dipoles. The fidelity can still be significantly degraded by a $p=1\,e$.nm dipole, but the map now shows clear trends. The worst dipole positions are beneath the left and right plunger gates (and, to a lower extent, below the $J$ gate) where the holes actually sit (see Fig.~\ref{fig:device}). In order to de-embed drive and dephasing errors, we plot in Fig.~\ref{fig:dipole}b the fidelity $\mathcal{F}_0^\mathrm{drive}$ of the drive calculated using Eq.~\eqref{eq:Fdrive} (the system is directly initialized in the operational states of the pristine device and projected back onto these states after the $\pi$-pulse). This drive fidelity (which discards phase information) is actually much better than $\mathcal{F}_0$ and mostly shows one major ``hot spot'' below the right plunger gate (which is just above the driven $R$ dot). Although the drive infidelity is sizable, the primary source of error in Fig.~\ref{fig:dipole}a is thus the dephasing accumulated throughout the full CNOT sequence. The hot spot below the left plunger gate is due to the fact that the elongated $L$ dot is highly polarizable along $x$. Note that such dephasing errors will not only limit two-qubit gates, but also single-qubit gates and idling qubits.

\begin{figure}[t]
    \centering
 	\includegraphics[width=\linewidth]{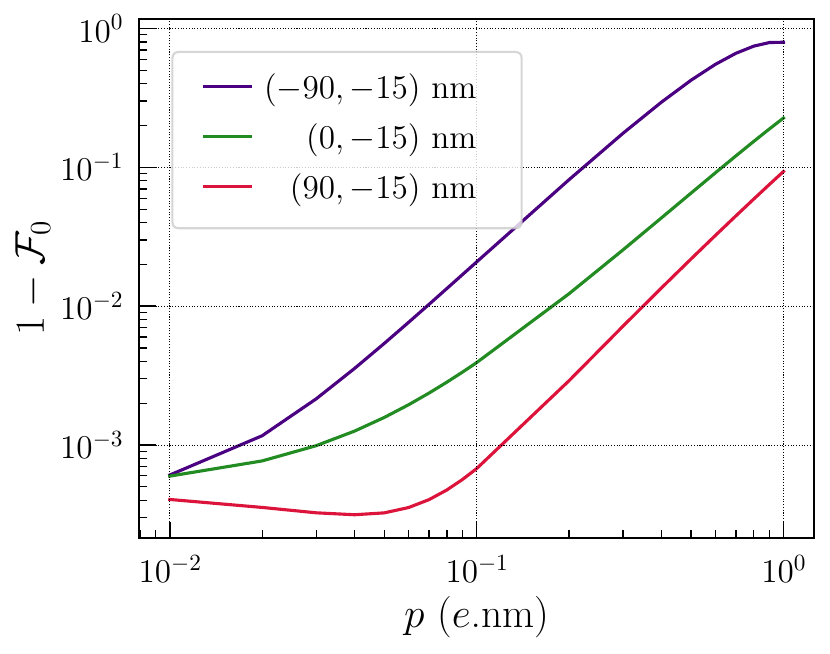}
    \caption{Infidelity $1-\mathcal{F}_0$ of the CNOT gate in the presence of a dipole $\vec{p}=p\vec{x}$ at the GeSi/Al$_2$O$_3$ interface as a function of its magnitude $p$ for three different positions in the $(xy)$ plane and $V_J^\mathrm{op}=-12.78\,$mV.}
    \label{fig:dipolestrengthF0}
\end{figure}

\begin{figure}[t]
    \centering
 	\includegraphics[width=\linewidth]{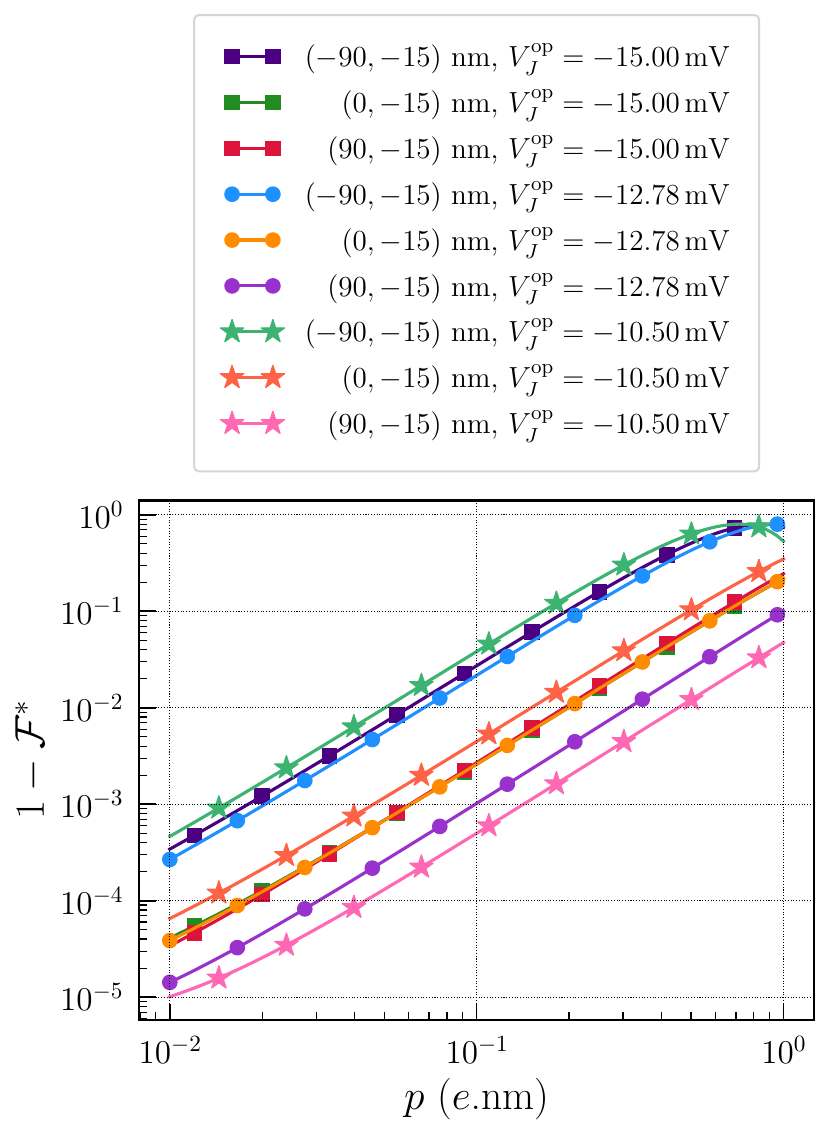}
    \caption{Infidelity $1-\mathcal{F}^*$ of the CNOT gate in the presence of a dipole $\vec{p}=p\vec{x}$ at the GeSi/Al$_2$O$_3$ interface as a function of its magnitude $p$ for three different positions in the $(xy)$ plane and three different $V_J^\mathrm{op}$.}
    \label{fig:dipolestrengthFs}
\end{figure}

We further plot in Figs.~\ref{fig:dipolestrengthF0} and \ref{fig:dipolestrengthFs} the infidelities $1-\mathcal{F}_0$ and $1-\mathcal{F}^*$ as a function of the strength $p$ of dipoles located at selected positions (below the plunger gates $P_L$, $P_R$ and the $J$-gate). As expected, $1-\mathcal{F}^*$, which characterizes the differences between the pristine and disordered device, steadily increases with increasing $p$ (approximately $\propto p^2$ in line with the lowest order Dyson expansion of the evolution operator \cite{Messiah61}). The behavior of $1-\mathcal{F}_0$ is more complex. Although it always increases for large dipoles (also $\propto p^2$), it can show a non monotonous behavior for small $p<0.1\,e$.nm. Indeed, the tiny effects of such a small dipole can compensate calibration errors and improve the net fidelity. 

There is no clear dependence of the $1-\mathcal{F}^*$ on the exchange energy $J_\parallel$, as illustrated by the data of Fig.~\ref{fig:dipolestrengthFs} for other $V_J^\mathrm{op}$. This is consistent with a fidelity primarily limited by the dephasing of the $L$ and $R$ dots (due to the modulation of their effective $g$-factors by the noise). 

\begin{figure}[t]
    \centering
 	\includegraphics[width=\linewidth]{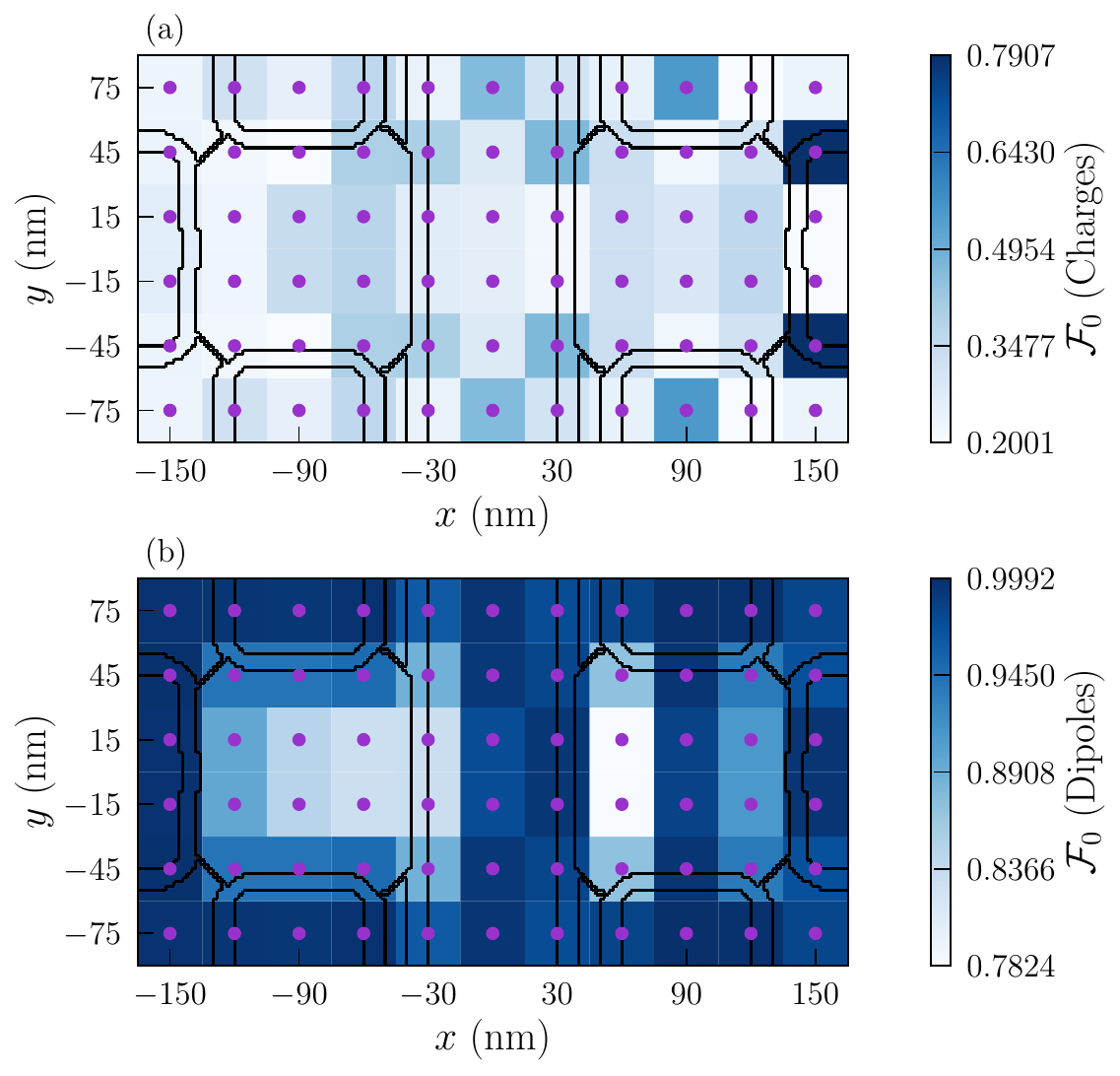}
    \caption{Fidelities $\mathcal{F}_0$ of the CNOT gate in the presence of (a) a charge $+e$ and (b) a dipole $\vec{p}=(1\,e\text{.nm})\vec{x}$ at the GeSi/Al$_2$O$_3$ interface, as a function of its position in the $(xy)$ plane (marked by the purple dots). The potentials of these charges and dipoles have been shifted $20$\,nm upward in order to mimic their effects in a $50$\,nm deep quantum well. The operational $J$-gate voltage is $V_J^\mathrm{op}=-12.78$\,mV where $J_\parallel=100$\,MHz. The position of the gates is delineated by the dark lines.}
    \label{fig:shiftedchargedipole}
\end{figure}

The top Ge$_{0.8}$Si$_{0.2}$ barrier of the present device is pretty thin ($t_\mathrm{GeSi}=30$\,nm) \cite{Hendrickx20}. It is, therefore, important to understand how the fidelity may be improved by a thicker barrier. We have, therefore, recomputed fidelities with the potentials of the charges and dipoles shifted $20$\,nm upward\footnote{Namely, the potentials are computed with the charges and dipoles at the GeSi/Al$_{2}$O$_{3}$ interface, then shifted by $20$\,nm along $z$ as if the well was buried $20$\,nm deeper.}. We mimic that way the disorder in a quantum well buried $50$\,nm below the gate stack, but can still compare the fidelity of exactly the same CNOT protocol as in the original device. The map of $\mathcal{F}_0$ for such shifted charge and dipole potentials is plotted in Fig.~\ref{fig:shiftedchargedipole}. For charges, the median fidelity over the map ($\approx 0.3$) is barely improved, even though the fidelity is much larger at some specific trap locations. This shows that the effects of charges are still very strong. Burying the well deeper would not, therefore, efficiently protect that DQD against charge traps. On the opposite, the fidelity is considerably better for dipoles. The map shows essentially the same structure as Fig.~\ref{fig:dipole}, but the maximum and median infidelities $1-\mathcal{F}_0$ are about four times smaller. In other words, a $50$\,nm deep well would withstand twice larger dipoles than the $30$\,nm deep well for the same loss of fidelity. The benefit is thus significant.

These results set constraints on the nature and strength of the noise permissible for high-fidelity two-qubit gates. They show, in particular, that the motion of carriers over $\gtrsim 1$\,\AA\ below the plunger and exchange gates is highly detrimental (unless a suitable sweet spot can be found for both qubits). This suggests that the quasi-static charge noise in faithful two-qubit gates is most likely resulting from remote traps.

\section{Conclusions}

To conclude, we have simulated a CNOT gate in a realistic layout for hole spin qubits in a Ge/GeSi heterostructure using a ``dressed basis set'' for the two-particle Hamiltonian. We have optimized the pulse timings in the pristine device and de-embedded the mechanisms limiting the fidelity of the CNOT operation. We find that the fidelity is mostly limited by the diabaticity of the exchange ramps, by the unwanted off-resonant driving of the complementary conditional transition, and to some extent by the AC Stark shifts, which complicate the calibration (as the drive frequencies become dependent on the drive power). The AC Stark shift is, however, highly dependent on the driving gate. Its effects can be almost completely suppressed (in the present layout) by driving with the $B_{RR}$ gate that exhibits a small Stark shift and is strongly coupled to the right dot. Off-resonant driving can also be strongly limiting but its impact can be largely mitigated by increasing the longitudinal exchange energy $J_\parallel$ (if possible) or by a careful choice of the drive amplitude (at the cost of a more complex calibration). The diabaticity of the exchange ramps is promoted by the transverse exchange interaction $J_\perp$, which calls for slower ramps for larger exchange energies. While enhancing exchange interactions suppresses off-resonant driving, it thus also slows down the exchange ramps. In the absence of noise (and with optimal drive amplitude), faster and more faithful CNOT gates can hence be achieved with moderate ($50-100$\,MHz) exchange energies than with larger ones.

We have next investigated the effects of quasi-static charge noise on the CNOT gate. The capture of a charge at the GeSi/Al$_2$O$_3$ interface in the vicinity of the qubits breaks the CNOT operation. The motion of a charge $q$ at this interface over a distance $d$ also degrades the fidelity according to its location and dipole $p=qd$. Locations below the plunger and exchange gates are, as expected, the most harmful. Achieving high fidelity ($>99.5$\%) CNOT gates is at best compatible with fluctuating dipoles in the $e$.\AA\ range around the qubits.

The framework introduced in Ref.~\cite{MRodriguez2025} and used in this work can be applied to a wide variety of materials and protocols, allowing for a better understanding of the physics and limitations of two-qubit gates, and opening the way for design optimizations. The conclusions on the diabaticity of the $J$-gate ramps and on the effects of charge noise apply, in particular, to the CPHASE and CZ gates \cite{Hendrickx21}, which are essentially a CNOT gate without drive shifting the phase of the $R$ qubit according to the state of the $L$ qubit.

\section*{Acknowlegements}

We thank Esteban A. Rodr\'iguez-Mena for fruitful discussions and suggestions. This work was supported by the ``France 2030'' program (PEPR PRESQUILE-ANR-22-PETQ-0002), and co‑funded by the European Union’s Chips Joint Undertaking (Chips JU) under grant agreement No. 101288553 SPINS.

\appendix

\section{Solution of the time-dependent Schr\"odinger equation for holes}
\label{app:TDSELK}

In this Appendix, we discuss the solution of the time-dependent Schr\"odinger equation for the Luttinger Kohn-Hamiltonian.

In the Luttinger-Kohn formalism \cite{Luttinger56}, the wave function of an electron in the valence band is expanded as a combination of heavy-hole (HH) and light-hole (HH) components mapped onto a $J=3/2$ angular momentum:
\begin{equation}
\psi_n(\vec{r},\sigma)=\sum_{\nu\in\left\{-\tfrac{3}{2},-\tfrac{1}{2},\tfrac{1}{2},\tfrac{3}{2}\right\}}\varphi_n^\nu(\vec{r})u_\nu(\vec{r},\sigma)\,,
\end{equation}
where $\sigma\in\{\uparrow,\downarrow\}$ is the physical spin, $u_\nu$ is the Bloch function with angular momentum $j_z=\nu$ along $z=[001]$ ($\nu=\pm 3/2$ for HH bands and $\nu=\pm 1/2$ for LH bands), and $\varphi_n^\nu$ is an envelope function.

The HH and LH envelope functions $\varphi_n^\nu$ in a magnetic field $\vec{B}$ (and vector potential $\vec{A}=\frac{1}{2}\vec{B}\times\vec{r}$) fulfill a set of differential equations defined by the Luttinger-Kohn Hamiltonian:
\begin{equation}
H_\mathrm{LK}(\vec{A},\vec{B})=-H_K-H_\varepsilon-H_Z-V_t(\vec{r})\mathbb{1}_4\,,
\label{eq:H4KP}
\end{equation}
where $H_K$ is the kinetic energy, $H_\varepsilon$ describes the effects of strains, $H_Z$ is the Zeeman Hamiltonian, $V_t(\vec{r})$ is the total potential and $\mathbb{1}_4$ is the $4\times4$ identity matrix. $H_K$ and $H_\varepsilon$ share the same generic form in the $J_z=\{+\tfrac{3}{2},+\tfrac{1}{2},-\tfrac{1}{2},-\tfrac{3}{2}\}$ basis set \cite{Luttinger56,Bir59,Winkler03,KP09}:
\begin{equation}
    H_{K/\varepsilon}=\begin{pmatrix}
      P+Q & -S & R & 0 \\
     -S^\dagger & P-Q & 0 & R \\
      R^\dagger & 0 & P-Q & S \\
      0 & R^\dagger & S^\dagger & P+Q
    \end{pmatrix}\,,
\label{eq:LK}
\end{equation}
where, for $H_K$,
\begin{subequations}
\begin{align}
    P_K&=\frac{1}{2m_0}\gamma_1(p_x^2+p_y^2+p_z^2) \\
    Q_K&=\frac{1}{2m_0}\gamma_2(p_x^2+p_y^2-2p_z^2) \\
    R_K&=\frac{1}{2m_0}\sqrt{3}\left[-\gamma_2(p_x^2-p_y^2)+2i\gamma_3\{p_x,\,p_y\}\right] \\
    S_K&=\frac{1}{2m_0}2\sqrt{3}\gamma_3\{p_x-ip_y,\,p_z\}\,,
\end{align}
\end{subequations}
with $\{U,\,V\}=\tfrac{1}{2}(UV+VU)$, and, for $H_\varepsilon$,
\begin{subequations}
\begin{align}
    P_\varepsilon&=-a_v(\varepsilon_{xx}+\varepsilon_{yy}+\varepsilon_{zz}) \\
    Q_\varepsilon&=-\frac{1}{2}b_v(\varepsilon_{xx}+\varepsilon_{yy}-2\varepsilon_{zz}) \\
    R_\varepsilon&=\frac{\sqrt{3}}{2}b_v(\varepsilon_{xx}-\varepsilon_{yy})-id_v\varepsilon_{xy} \\
    S_\varepsilon&=-d_v(\varepsilon_{xz}-i\varepsilon_{yz})\,.
\end{align}
\end{subequations}
Here $\vec{p}$ is the momentum, $m_0$ is the free electron mass, and $\gamma_1$, $\gamma_2$, $\gamma_3$ are the Luttinger parameters that characterize the hole masses. The $\varepsilon_{\alpha\beta}$ are the strains; $a_v$ is the hydrostatic, $b_v$ the uniaxial and $d_v$ the shear deformation potential of the valence band. The Zeeman Hamiltonian
\begin{equation}
    H_Z=2\mu_B(\kappa\vec{B}\cdot\vec{J}+q\vec{B}\cdot\vec{J}^3)
    \label{eq:HZ}
\end{equation}
describes the action of the magnetic field on the Bloch functions, with $\vec{J}$ the spin $\tfrac{3}{2}$ operator, $\vec{J}^3\equiv(J_x^3,J_y^3,J_z^3)$, $\mu_B$ the Bohr magneton, and $\kappa$, $q$ the isotropic and cubic Zeeman parameters. The action of $\vec{B}$ on the envelope functions is accounted for by the substitution $\vec{p}\to-i\hbar\boldsymbol{\nabla}+e\vec{A}$ in $H_K$. Material parameters can be found in Ref.~\cite{Abadillo2023}.

Equation~\eqref{eq:H4KP} is a Hamiltonian for electrons. In a given basis set $\{\ket{\chi_n}\}$ of single-particle orbitals, it can be second-quantized as:
\begin{equation}
    H_\mathrm{LK}=\sum_{i,j} h_{ij}c_i^\dagger c_j\text{ with }h_{ij}=\bra{\chi_i}H_\mathrm{LK}\ket{\chi_j}\,,
    \label{eq:2ndhole1}
\end{equation}
where $c_i^\dagger$ is the operator creating an electron in state $\ket{\chi_i}$. Introducing the hole creation operator $b_i^\dagger=c_i$ and annihilation operator $b_i=c_i^\dagger$,
\begin{equation}
    H_\mathrm{LK}={\cal E}_0-\sum_{i,j} h_{ij}^*b_i^\dagger b_j\,,
\end{equation}
where ${\cal E}_0$ is the energy of the vacuum of holes (the energy of the $N$-electron ground-state). Leaving aside this constant,
\begin{equation}
    H_\mathrm{LK}=\sum_{i,j} {\tilde h}_{ij}b_i^\dagger b_j
\end{equation}
with:
\begin{equation}
    {\tilde h}_{ij}=-h_{ij}^*=\bra{\chi_i^*}(-H_\mathrm{LK}^*)\ket{\chi_j^*}\,.
    \label{eq:2ndhole2}
\end{equation}
In other words, the hole energies and wave functions are the eigenvalues and eigenvectors of $-H_\mathrm{LK}^*$ (hence the opposite of the electron energies and the complex conjugates of the electron wave functions) \cite{Winkler03}. 

Quite generally, if $H_\mathrm{elec}$ is a single or many-electron Hamiltonian matrix, the corresponding hole Hamiltonian matrix is $H_\mathrm{hole}=-H_\mathrm{elec}^*=-H_\mathrm{elec}^T$, and the time-dependent Schr\"odinger equation for the hole state $\ket{\psi_\mathrm{hole}}$ reads
\begin{equation}
    i\hbar\frac{\partial}{\partial t}\ket{\psi_\mathrm{hole}}=-H_\mathrm{elec}^*\ket{\psi_\mathrm{hole}}\,,
    \label{eq:TDSEholes}
\end{equation}
whose solution is, as expected, the conjugate of the electron state $\ket{\psi_\mathrm{elec}}$ that satisfies $i\hbar\partial_t\ket{\psi_\mathrm{elec}}=H_\mathrm{elec}\ket{\psi_\mathrm{elec}}$. In this work, we actually use $H_\mathrm{hole}=-H_\mathrm{elec}^*$ as hole Hamiltonian and thus solve Eq.~\eqref{eq:TDSEholes} for the dynamics of the CNOT gate.

Nevertheless, it is customary to work with $-H_\mathrm{LK}$ as hole Hamiltonian and forget about the complex conjugation (as done in the original LK paper \cite{Luttinger56} or in our previous work \cite{MRodriguez2025}). We analyse below the implications of this practice.


There is {\it a priori} no obvious relation between the solutions of $i\hbar\partial_t\ket{\psi}=-H_\mathrm{elec}^*\ket{\psi}$ and $i\hbar\partial_t\ket{\psi}=-H_\mathrm{elec}\ket{\psi}$. For the LK Hamiltonian, complex conjugation of Eq.~\eqref{eq:LK} indeed changes $R$ into $R^\dagger$, $S$ into $S^\dagger$, and Peierls-Onsager substitution $\vec{p}\to-i\hbar\boldsymbol{\nabla}+e\vec{A}$ into\footnote{Actually $\vec{p}\to i\hbar\boldsymbol{\nabla}+e\vec{A}$, but the LK Hamiltonian is invariant under the transformation $\vec{p}\to -\vec{p}$.} $\vec{p}\to-i\hbar\boldsymbol{\nabla}-e\vec{A}$ (as actually expected for holes). It also leaves $J_x$ and $J_z$ invariant\footnote{The matrices of $J_x$, $J_y$ and $J_z$ consistent with Eq.~\eqref{eq:LK} are:
\begin{gather}
    J_x=\frac{1}{2}\begin{pmatrix}
        0 & \sqrt{3} & 0 & 0 \\
        \sqrt{3} & 0 & 2 & 0 \\
        0 & 2 & 0 & \sqrt{3} \\
        0 & 0 & \sqrt{3} & 0
    \end{pmatrix}\,, \\
    J_y=\frac{i}{2}\begin{pmatrix}
        0 & -\sqrt{3} & 0 & 0 \\
        \sqrt{3} & 0 & -2 & 0 \\
        0 & 2 & 0 & -\sqrt{3} \\
        0 & 0 & \sqrt{3} & 0
    \end{pmatrix}\,,\,
    J_z=\frac{1}{2}\begin{pmatrix}
        3 & 0 & 0 & 0 \\
        0 & 1 & 0 & 0 \\
        0 & 0 & -1 & 0 \\
        0 & 0 & 0 & -3
    \end{pmatrix}\,. \nonumber 
\end{gather}} in $H_Z$ [Eq.~\eqref{eq:HZ}], but changes $J_y$ into $-J_y$. Nevertheless, the unitary transformation of the Bloch functions ${\tilde H}_\mathrm{LZ}=P^\dagger H_\mathrm{LZ}^*P$ with
\begin{equation}
    P=\begin{pmatrix}
 		0 & 0 & 0 & 1 \\
 		0 & 0 & -1 & 0 \\
 		0 & 1 & 0 & 0 \\
 		-1 & 0 & 0 & 0
	\end{pmatrix}
\end{equation}
brings back $H_K$ and $H_\varepsilon$ to their original forms, but changes $H_Z$ into $-H_Z$ (or equivalently $\vec{B}$ into $-\vec{B}$). Since the modified Peierls-Onsager substitution $\vec{p}\to-i\hbar\boldsymbol{\nabla}-e\vec{A}$ also amounts to change $\vec{A}$ into $-\vec{A}$, we conclude that $-H_\mathrm{LK}^*(\vec{A},\vec{B})$ and $-H_\mathrm{LK}(-\vec{A},-\vec{B})$ describe the same physics\footnote{However, making the substitution $\vec{p}\to-i\hbar\boldsymbol{\nabla}-e\vec{A}$ in $-H_\mathrm{LK}$ without changing the sign of $H_Z$ is incorrect.}. 

Therefore, using $-H_\mathrm{LK}(\vec{A},\vec{B})$ [instead of $-H_\mathrm{LK}^*(\vec{A},\vec{B})$] as hole Hamiltonian provides solutions for the opposite magnetic field $(-\vec{A},-\vec{B})$. If the quantities of interest are even with respect to the magnetic field, the outcome is the same. This is actually the case in this work, and we have verified that solving the time-dependent Schr\"odinger equation with either $-H_\mathrm{elec}$ or $-H_\mathrm{elec}^*$ does not change the observables such as the fidelity (although the wave functions are different).

Practically, the two-particle Hamiltonian drawn from configuration interaction with the single-particle LK wave functions is projected onto a small ``dressed'' basis set of low-energy singlet and triplet states sampled on a grid of relevant bias points (see details in Ref.~\cite{MRodriguez2025}). The time-dependent Schr\"odinger equation is then solved on a grid of times $t=t_0,\,...,\,t_N$; for that purpose, the evolution operator $U(t_{n+1},\,t_n)=\exp[iH_\mathrm{hole}(t_{n+1/2})(t_n-t_{n+1})/\hbar]$ is computed in each time interval $[t_n,\,t_{n+1}]$ from the exact diagonalization of the Hamiltonian $H_\mathrm{hole}(t_{n+1/2})$ [with $t_{n+1/2}=(t_n+t_{n+1})/2$]. The control signals are, therefore, assumed constant over each time interval.

\section{Expressions for $J_0$, $J_\parallel$ and $J_\perp$ near the symmetric operation point}
\label{app:HundMulliken}

To derive Eqs.~\eqref{eq:Es2}, we start from the Hamiltonian in the singlet/triplet basis set $\{S(2,0),S(0,2),S(1,1),T_0(1,1),T_+(1,1),T_-(1,1)\}$ built upon the Zeeman-split states $\{\ket{L\Uparrow},\ket{L\Downarrow},\ket{R\Uparrow}\ket{R\Downarrow}\}$ of each dot [Eq.~(21) of Ref.~\cite{MRodriguez2025}]:
\begin{equation}
    H=\begin{bmatrix}    
        U_L+\varepsilon & 0 & \sqrt{2}t_\mathrm{sc} & 0 & -t_\mathrm{sf}^* & -t_\mathrm{sf} \\
        0 & U_R-\varepsilon & \sqrt{2}t_\mathrm{sc} & 0 & -t_\mathrm{sf}^* & -t_\mathrm{sf} \\
        \sqrt{2}t_\mathrm{sc} & \sqrt{2}t_\mathrm{sc} & 0 & \frac{1}{2}\Delta E_Z & 0 & 0 \\
        0 & 0 & \frac{1}{2}\Delta E_Z & 0 & 0 & 0 \\
        -t_\mathrm{sf} & -t_\mathrm{sf} & 0 & 0 & \bar{E}_Z & 0 \\
        -t_\mathrm{sf}^* & -t_\mathrm{sf}^* & 0 & 0 & 0 & -\bar{E}_Z
    \end{bmatrix}\,.
    \label{eq:6bandH}
\end{equation}
Here $U_L$ and $U_R$ are the charging energies of the dots, $\varepsilon=\varepsilon_L-\varepsilon_R$ is the detuning between their single-particle ground-state energies $\varepsilon_L$ and $\varepsilon_R$, $\bar{E}_Z=\mu_B(g_L^*+g_R^*)B/2$ and $\Delta E_Z=\mu_B(g_L^*-g_R^*)B$ are the average and difference between their Zeeman energies (with $g_L^*$, $g_R^*$ their effective $g$-factors), and $t_\mathrm{sc}$, $t_\mathrm{sf}$ are the spin-conserving and spin-flip tunnel couplings. Note that $\bar{E}_Z$, $\Delta E_Z$, $t_\mathrm{sc}$ and $t_\mathrm{sf}$ depend on the orientation of the magnetic field, but that the net tunnel coupling $t=\sqrt{t_\mathrm{sc}^2+\abs{t_\mathrm{sf}}^2}$ does not (in the absence of magneto-tunneling effects \cite{rodriguezmena2025}).

\begin{figure}
	\centering
  	\includegraphics[width=\linewidth]{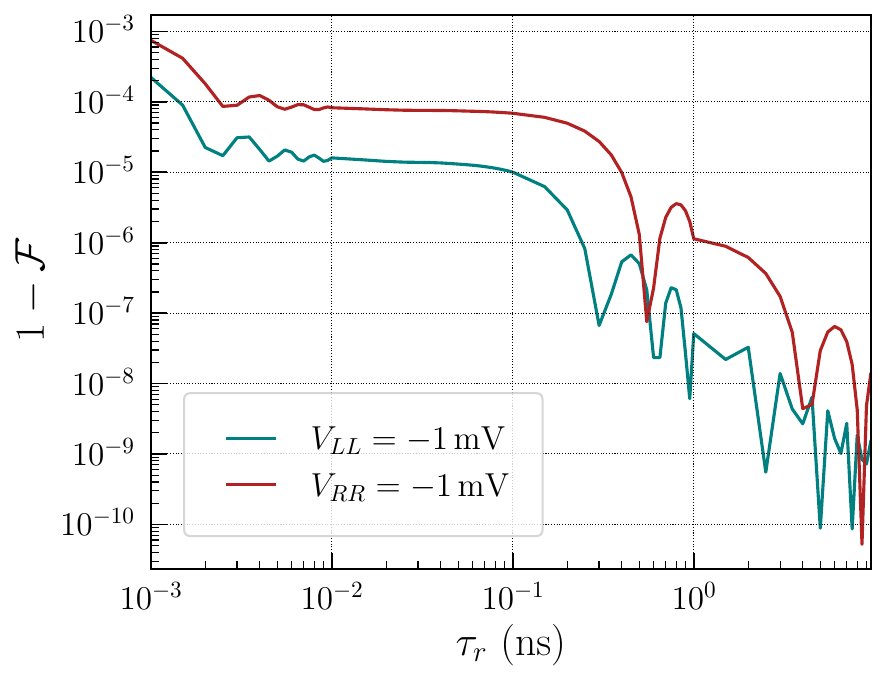}
	\caption{(a) Infidelity of the phase gate ramps as a function of the ramp time $\tau_r$.}
	\label{fig:Rramp}
\end{figure}

We next assume $\varepsilon\ll U_L,U_R$ (DQD biased close to the symmetric operation point), and decouple the $S(2,0)$ and $S(0,2)$ states by a Schrieffer-Wolff transformation to get an effective Hamiltonian $H^\mathrm{SW}$ in the $\{S(1,1),T_0(1,1),T_+(1,1),T_-(1,1)\}$ subspace:
\begin{equation}
H_{ij}^\mathrm{SW}=H_{i+2,j+2}+\frac{H_{i+2,1}H_{1,j+2}}{U_L+\varepsilon}+\frac{H_{i+2,2}H_{2,j+2}}{U_R-\varepsilon}\,.
\end{equation}
We finally rotate this Hamiltonian in the $\{\ket{\Uparrow\Uparrow},\ket{\Uparrow\Downarrow},\ket{\Downarrow\Uparrow},\ket{\Downarrow\Downarrow}\}$ basis set as $H_e=R^\dagger H^\mathrm{SW}R$ with
\begin{equation}
    R=\begin{pmatrix}
 		0 & 0 & 1 & 0 \\
 		\frac{1}{\sqrt{2}} & \frac{1}{\sqrt{2}} & 0 & 0 \\
 		-\frac{1}{\sqrt{2}} & \frac{1}{\sqrt{2}} & 0 & 0 \\
 		0 & 0 & 0 & 1
	\end{pmatrix}\,.
\end{equation}
This yields (up to a global shift):
\begin{equation}
    H_e=\begin{pmatrix}
        \bar{E}_Z & \gamma t_\mathrm{sc}t_\mathrm{sf} & -\gamma t_\mathrm{sc}t_\mathrm{sf} & -\gamma t_\mathrm{sf}^2\\
        \gamma t_\mathrm{sc}t_\mathrm{sf}^* & -\frac{J_\parallel}{2}+\frac{\Delta E_Z}{2} & \frac{J_\perp}{2} & \gamma t_\mathrm{sc}t_\mathrm{sf}\\
        -\gamma t_\mathrm{sc}t_\mathrm{sf}^* & \frac{J_\perp}{2} & -\frac{J_\parallel}{2}-\frac{\Delta E_Z}{2} & -\gamma t_\mathrm{sc}t_\mathrm{sf}\\
        -\gamma t_\mathrm{sf}^{*2} & \gamma t_\mathrm{sc}t_\mathrm{sf}^* & -\gamma t_\mathrm{sc}t_\mathrm{sf}^* & -\bar{E}_Z
    \end{pmatrix}\,,
    \label{eq:Heff2}
\end{equation}
where:
\begin{subequations}
\begin{align}
    \gamma &=\frac{1}{U_L+\varepsilon}+\frac{1}{U_R-\varepsilon}\\
    J_\parallel&=2\gamma(t_\mathrm{sc}^2-\abs{t_\mathrm{sf}}^2)\\
    J_\perp&=2\gamma t_\mathrm{sc}^2
\end{align}
\end{subequations}
At high enough magnetic field such that $\bar{E}_Z\gg\abs{\Delta E_Z}$, we may neglect the off-diagonal, $\propto\gamma$ terms in Eq.~\eqref{eq:Heff2} and decouple that way the $\{\ket{\Uparrow\Uparrow},\ket{\Downarrow\Downarrow}\}$ from the $\{\ket{\Uparrow\Downarrow},\ket{\Downarrow\Uparrow}\}$ subspace. We then fall back to Eq.~\eqref{eq:Heff} of the main text. On the opposite, when $B\to 0$ the eigenstates of Eq.~\eqref{eq:Heff2} are a singlet split from a triplet by an energy:
\begin{equation}
J_0=2\gamma(t_\mathrm{sc}^2+\abs{t_\mathrm{sf}}^2)\,.
\end{equation}
The different exchange energies thus fulfill $J_\perp=(J_0+J_\parallel)/2$, and we recover Eqs.~\eqref{eq:Es2} by settting $U_L=U_R=U_0$.

\section{Fidelity of the phase gate ramps}
\label{app:phasegateramps}

We plot in Fig.~\ref{fig:Rramp} the infidelities of the ramps of the two phase gates ($V_{LL}=-1$\,mV and $V_{RR}=-1$\,mV) as a function of the ramp time $\tau_r$. Given the small voltage span, the ramps enter in the quasi-adiabatic regime once $\tau_r\gtrsim 0.5$\,ns, and the infidelity hardly exceeds $10^{-4}$ down to picosecond ramp times. The ramp is slightly less faithful on the $B_{RR}$ gate than on the $B_{LL}$ gate.

\section{Time traces for combined $J$-gate ramps and $\pi$-pulse}
\label{app:timetraces}

\begin{figure}
	\centering
    \includegraphics[width=\linewidth]{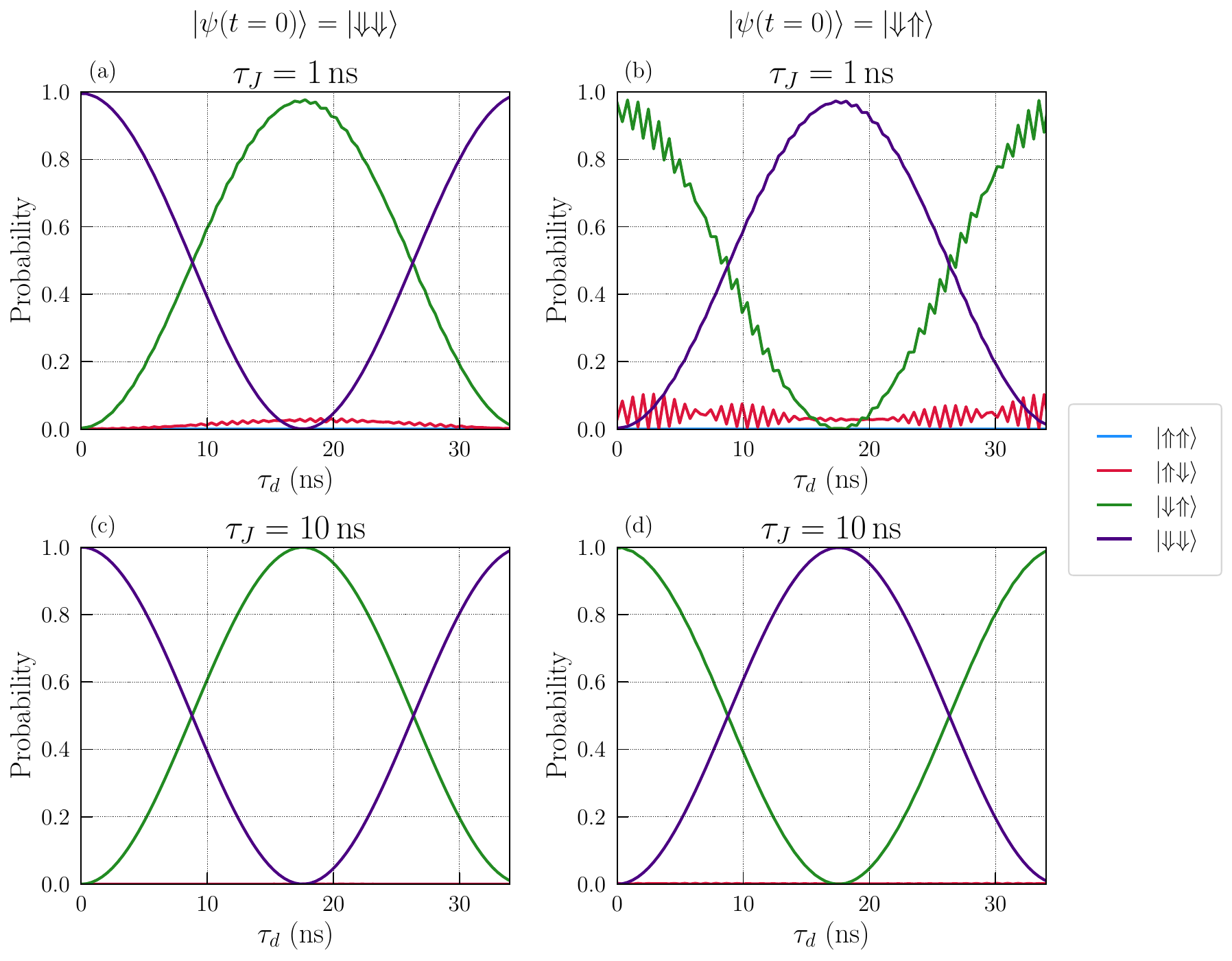}
	\caption{Probability of being in the computational basis states $\{\ket{\Uparrow\Uparrow},\ket{\Uparrow\Downarrow},\ket{\Downarrow\Uparrow},\ket{\Downarrow\Downarrow}\}$ as a function of the driving time $\tau_d$ for different $J$-gate ramp times $\tau_J$ ($V_J^\mathrm{op}=-15$\,mV). In panels (a, b), $\tau_J=1$\,ns (diabatic ramp), and in panels (c, d), $\tau_J=10$\,ns (adiabatic ramp). The initial state is $\ket{\Downarrow\Downarrow}$ in panels (a, c) and $\ket{\Downarrow\Uparrow}$ in panels (b, d).}
	\label{fig:traces}
\end{figure}

We plot in Fig.~\ref{fig:traces} the probability to be in the computational basis states $\{\ket{\Uparrow\Uparrow},\ket{\Uparrow\Downarrow},\ket{\Downarrow\Uparrow},\ket{\Downarrow\Downarrow}\}$ as a function of the drive time $\tau_d$, for two $J$-gate ramp times $\tau_J=1$\,ns and $\tau_J=10$\,ns ($V_J^\mathrm{op}=-15$\,mV), and two initial states $\ket{\psi(t=0)}=\ket{\Downarrow\Downarrow}$ and $\ket{\psi(t=0)}=\ket{\Downarrow\Uparrow}$. Starting from any of these initial states, the barrier is open over time $\tau_J$, the DQD is driven at frequency $f_\Downarrow$ for time $\tau_d$ with gate $B_{RR}$, the barrier is closed again over time $\tau_J$, and the final states are projected on the computational basis states.

When $\tau_J=10$\,ns, the $J$-gate ramps are adiabatic and the computational basis states $\{\ket{\Uparrow\Uparrow},\ket{\Uparrow\Downarrow},\ket{\Downarrow\Uparrow},\ket{\Downarrow\Downarrow}\}$ map onto the operational basis states $\{\ket{3},\ket{2},\ket{1},\ket{0}\}$. The DQD thus exhibits the expected, high visibility Rabi oscillations between $\ket{\Downarrow\Downarrow}\equiv\ket{0}$ and $\ket{\Downarrow\Uparrow}\equiv\ket{1}$. However, the $J$-gate ramps are not adiabatic any more at small $\tau_J=1$\,ns. The initial $\ket{\Downarrow\Downarrow}$ state is still reasonably transformed into $\ket{0}$ by the opening ramp, but the $\ket{1}$ component of the driven state is transformed into a combination of a major $\ket{\Downarrow\Uparrow}$ and a minor $\ket{\Uparrow\Downarrow}$ component by the closing ramp owing to the finite transverse exchange $J_\perp$. As a consequence, the probability to end in $\ket{\Uparrow\Downarrow}$ shows spurious oscillations synchronous with the probability to be in the $\ket{\Downarrow\Uparrow}$ state. The situation gets even worse starting from the $\ket{\Downarrow\Uparrow}$ state. Indeed, the latter is transformed into a combination of the operational $\ket{1}$ and $\ket{2}$ states by the opening ramp. There are, therefore, fast $\ket{1}\leftrightarrow\ket{2}$ oscillations developing over the driven $\ket{0}\leftrightarrow\ket{1}$ transition, well visible in the $\ket{\Downarrow\Uparrow}$ and $\ket{\Uparrow\Downarrow}$ probabilities after the closing ramp. Both processes contribute to the degradation of the net fidelity of the combined $J$-gate ramps and drive.

\vfill 

\bibliography{Refs.bib}

\end{document}